\documentclass[
	aps, prd,  notitlepage, 
        floats, floatfix, onecolumn,
	amsmath, amssymb, amsfonts, eqsecnum,
	superscriptaddress,
	showpacs, showkeys,
	nofootinbib,
 	longbibliography,
]{revtex4-1}
\usepackage{multirow}
\usepackage{graphicx}
\usepackage{subfigure}
\usepackage[usenames,dvipsnames]{xcolor}
\usepackage{tikz}
\usepackage{xspace} % Sensible space treatment at end of simple macros
\usepackage{bm}% bold math
\usepackage[utf8]{inputenc} % for some references from inspirehep.net
\usepackage{latexsym}
\usepackage{float}
\usepackage{ulem}
\renewcommand{\emph}[1]{\textit{#1}}
\xdefinecolor{mylinkcolor}{rgb}{0,0,0.5}
\usepackage[
	bookmarksnumbered, bookmarksopen, bookmarksopenlevel=2,
	breaklinks=true,
	colorlinks=true, filecolor=mylinkcolor, citecolor=mylinkcolor,
	linkcolor=mylinkcolor, urlcolor=mylinkcolor, menucolor=mylinkcolor,
]{hyperref}

\def\be{\begin{equation}}
\def\ee{\end{equation}}
\def\bea{\begin{eqnarray}}
\def\eea{\end{eqnarray}}
\newcommand{\bes}{\begin{subequations}}
\newcommand{\ees}{\end{subequations}}
\def\comment#1{}

\allowdisplaybreaks

\begin{document}

\title{Analytical effective-one-body formalism for extreme-mass-ratio inspirals: eccentric orbits}
\date{\today}

\author{Chen Zhang}
\author{Wen-Biao Han}
\email{wbhan@shao.ac.cn}
\author{Shu-Cheng Yang}
\affiliation{Shanghai Astronomical Observatory, Chinese Academy of Sciences, Shanghai 200030, P. R. China}
\affiliation{School of Astronomy and Space Science, University of Chinese Academy of Sciences Beijing, 100049, P. R. China}

\begin{abstract}
Extreme-mass-ratio-inspiral (EMRI) is one of the most important sources for the future space-borne gravitational wave detectors. In such kind of systems, the compact objects usually orbit around the central supermassive black holes with complicated trajectories. Usually, the trajectory is approximated as geodesic of a test-particle in Kerr space-time, and the orbital evolution are simulated with the help of adiabatic approximation. However, this omits the influence of the compact object on the back ground. In the present paper, employing effective-one-body formalism, we analytically calculate out the trajectories of a compact object around a massive Kerr black hole in equatorial-eccentric orbit, and express the fundamental orbital frequencies in explicit forms. Our formalism include the first-order corrections of mass-ratio in the conservative orbital motion. Furthermore, we insert the mass-ratio related terms in the first post-Newtonian energy fluxes. By calculating the gravitational waves from the Teukolsky equations, we quantitatively reveal the influence of the mass of the compact object on the data analysis. We find that the shrinking of geodesic motion by taking the small objects as test particles may be not appropriate for the detection of EMRIs. 
\end{abstract}

\maketitle

\section{Introduction}
The successful detection of gravitational waves (GWs) by Advanced LIGO and Virgo~\cite{abbott2016observation,abbott2016gw151226,scientific2017gw170104,abbott2017gw170608,abbott2017gw170814,abbott2017gw170817} announces that the era of GW Astronomy is coming. This kind of ground-based detectors observe the GWs in high frequency band. LISA, a space-borne gravitational wave (GW) detector which proposed by Europe and USA~\cite{danzmann1996lisa}, at the same time, two Chinese space projects Taiji~\cite{hu2017taiji} and Tian-Qin~\cite{luo2016tianqin}, will be planed to launch after 2030. All these detectors focus on GWs at low frequency (about 0.1 mHz to 1 Hz). Extreme-mass-ratio inspirals (EMRIs) composed by compact objects (stellar black holes, neutron stars, white drawfs and etc.) and supermassive black holes (SMBHs), are expected as one of the most important sources for these space-borne detectors~\cite{amaro2007intermediate,babak2017science,berry2019unique}.

The signals from EMRIs usually are very weak, but with one years' observation, the signal-to-noise ratio can be enough to be detected by matched filtering technology~\cite{amaro2007intermediate}. For detecting this kind of long duration signals, the requirement of accuracy of waveform templates is very high. Typically, after $10^5$ cycles, the dephaseing should be less than few radians~\cite{gair2013testing,babak2017science}. Nowadays, there are a few of EMRI templates, like as AK~\cite{barack2004lisa}, AAK~\cite{chua2017augmented}, NK~\cite{babak2007kludge}, XSPEG~\cite{xin2019gravitational} and so on. All of them take the small object as a test particle and omit the mass in their conservation dynamics part. Some works considered the correction due the small mass by using effective-one-body (EOB) formalism, but only in circular orbits~\cite{yunes2010modeling,yunes2011extreme} or for eccentric orbits with data fitted parameters~\cite{han2014gravitational}. There are also intermediate-mass-ratio inspirals (IMRIs) composed with stellar compact objects and intermediate massive black holes(IMBHs) or IMBHs orbiting SMBHs, and the mass-ratio of IMRI is around $10^{-3}$~\cite{amaro2018detecting}. In this situation, the mass-ratio correction on the conservative orbital dynamics should be more important. 

The EOB formalism, by including the mass-ratio corrections in post-Newtonian (PN) expansions, can well describe the dynamical evolution of binary black holes~\cite{buonanno1999eff,buonanno2000transition}, and are widely used to construct the waveform templates for LIGO~\cite{taracchini2014effective,buonanno2007approaching,purrer2016frequency,husa2016frequency,khan2016frequency,chu2016accuracy,kumar2016accuracy,pan2014inspiral}. Most of these models only considered the circular orbit cases. Recently, Hinderer et. al. gave an analytical eccentric EOB dynamics for Schwarzschild BHs~\cite{hinderer2017foundations}. Cao and Han built an eccentric EOBNR waveform template (SEOBNRE) for spinning black holes~\cite{cao2017waveform}, but the orbits did not be geometrized and the orbital parameters did not be well defined. 

It is well known that the orbits of EMRIs could be highly eccentric~\cite{babak2017science}, and the supermassive black hole in the center should be spinning in general. In the present paper, we extend the previous work by Hinderer and Babak to the Kerr black holes. As a start, for equatorial-eccentric EMRIs, we analytically transfer the original EOB dynamical equations to geometric kinetic motion with orbital parameters: semilatus rectum $p$ and the eccentricity $e$ together with two phase variables associated with the spatial geometry of the radial and azimuthal motion denoted by $(\xi,\phi)$. Because of the extreme small mass-ratio, we omit the spin of the effective small body, then no the very complicated spin-spin coupling terms. 

An important feature of the dynamics of an extreme-mass-ratio binary system on a bounded equatorial-eccentric orbit is that the orbit can be characterized by two frequencies: the radial frequency $\omega_r$ associated with the libration between the apo- and periapsis, and the azimuthal rotational frequency $\omega_\phi$. Once these two frequencies and orbital parameters are obtained, one can solve the Teukolsky equations~\cite{Teukolsky} to get the accurate waveforms of the eccentric EMRIs. The combination of EOB and Teukolsky-based waveforms has been implemented by one of the authors, and was called as ET codes~\cite{han2010gravitational,han2011constructing,han2014gravitational,han2017excitation,cai2016gravitational,yang2019testing,cheng2019highly}. 

The organization of this paper is as follows. In Sec.~\ref{sec:geometrization}, we re-parameterize the origial spinning EOB dynamical  description to a geometric formalism in the more efficient re-parameterized terms of $(p,e,\xi,\phi)$. We analytical express the fundamental frequencies in two integrals with parameter $\xi$. Next, we focus on the evolution of orbital parameters with gravitational radiation reaction with PN fluxes. We also show the waveforms calculated from the Teukolsky equations. Especially, we investigate the influence of mass-ratio on the detection of EMRIs. Section~\ref{sec:fluxes} contains our conclusions and the outlook on remaining tasks for future work. Finally, the Appendices contain details about the EOB formalism and the expressions of orbital evolution in details. 

Throughout this paper we will use geometric units $G = c = 1$, and the units of time and length is the mass of system $M$, and the unit of linear and angular momentum are $\mu$ and $\mu M$ respectively, where $\mu$ is the reduced mass of the effective body.

%%%%%%%%
\section{Geometrization of the conservative dynamics in deformed Kerr spacetime}
\label{sec:geometrization}
%%%%%%%%
\subsection{The effective-one-body Hamiltonian}
\label{sec:HEOB}
%%%%%%%%
 The EOB formalism was originally introduced in~\cite{buonanno1999eff,buonanno2000transition} to describe the evolution of binary system. We start by considering  an EMRI system with central Kerr black hole $m_1$ and inspiraling object $m_2$ (assume it is nonspinning for simplicity) which is restricted on the equatorial plane of $m_1$ ($m_2 \ll m_1$). For the moment, we neglect any radiation reaction effects and focus on purely geodesic motion. The conservative orbital dynamics is derived via Hamilton’s equations using the EOB Hamiltonian $H_{\rm EOB}=M\sqrt{1+2\nu (\hat{H}_{\rm eff}-1)}$, where $M=m_1+m_2$, $\nu=m_1m_2/M$ , $\mu=\nu M$, and $\hat{H}_{\rm eff}=H_{\rm eff}/\mu$ . The  deformed-Kerr metric is given by~\cite{barausse2010improved}
\begin{subequations}
\begin{eqnarray}
\label{def_metric_in}
g^{tt} &=& -\frac{\Lambda_t}{\Delta_t\,\Sigma}\,,\\
g^{rr} &=& \frac{\Delta_r}{\Sigma}\,,\\
g^{\theta\theta} &=& \frac{1}{\Sigma}\,,\\
g^{\phi\phi} &=& \frac{1}{\Lambda_t}
\left(-\frac{\widetilde{\omega}_{\rm fd}^2}{\Delta_t\,\Sigma}+\Sigma\right)\,,\\
g^{t\phi}&=&-\frac{\widetilde{\omega}_{\rm fd}}{\Delta_t\,\Sigma}\,,\label{def_metric_fin}
\end{eqnarray}
\end{subequations}
The quantities $\Sigma$, $\Delta_t$, $\Delta_r$, $\Lambda_t$ and $\widetilde{\omega}_{\rm fd}$
in Eqs.~(\ref{def_metric_in})--(\ref{def_metric_fin}) are given by 
\begin{eqnarray}
\Sigma &=& r^2 \,, \\
\label{deltat}
\Delta_t &=& r^2\, \left [A(u) + \frac{a^2}{M^2}\,u^2 \right ]\,, \\
\label{deltar}
\Delta_r &=& \Delta_t\, D^{-1}(u)\,,\\
\Lambda_t &=& (r^2+a^2)^2 - a^2\,\Delta_t \,,\\
\widetilde{\omega}_{\rm fd}&=& 2 a\, M\, r
+ \omega_1^{\rm fd}\,\nu\,
\frac{a M^3}{r}
+ \omega_2^{\rm fd}\,\nu\,
\frac{M a^3}{r}
\label{eq:omegaTilde}\,,
\end{eqnarray}
where $a$ is the effective Kerr parameter and $u = M/r$. The values of $\omega_1^{\rm fd}$ and $\omega_2^{\rm fd}$ given by a preliminary comparison of EOB model with numerical relativity results are about $-10$ and $20$, respectively. The metric potentials $A$ and $D$ for the EOB model are given in the Appendix~\ref{sec:potential}. The effective Hamiltonian associated with the metric~(\ref{def_metric_in})--(\ref{def_metric_fin}) has the form~\cite{barausse2010improved}
\begin{equation}
H_{\rm eff} = H_{\rm NS} + H_{\rm S}\,,
\end{equation}
where $H_{\rm S}$ is the Hamiltonian caused by the spin of the effective particle. Considering the effective spin $s \sim \mu a/M$ is very small for EMRIs, we omit this term in the present work for simplification and will include it in the next work. The most dominant part, ${H}_{\rm NS}$ is the Hamiltonian for a non-spinning test-particle of mass $\mu$, given by
\begin{equation}\label{eq:Hns}
{H}_{\rm NS} = \beta^iP_i + \alpha \sqrt{\mu^2 + \gamma^{ij}\,P_i\,P_j}\,,
\end{equation}
with
\begin{eqnarray}
\label{alphai}
\alpha &=& \frac{1}{\sqrt{-g^{tt}}}\,,\\
\beta^i &=& \frac{g^{ti}}{g^{tt}}\,,\\
\gamma^{ij} &=& g^{ij}-\frac{g^{ti} g^{tj}}{g^{tt}}\,,
\label{gammai}
\end{eqnarray}

The energy of the system is given by
\be
{ E}=H_{\rm EOB},
\ee
which implies the relation
\be
\hat{H}_{\rm eff}(E)=1+\frac{1}{2\nu}\left(\frac{E^2}{M^2}-1\right). \label{eq:HeffofE}
\ee
the canonical EOB dynamics without radiation reaction
\be
%\left\{
\begin{aligned}
\dot r & =\frac{\partial H_{\rm EOB}}{\partial P_r}, \dot P_r=-\frac{\partial H_{\rm EOB}}{\partial r}   \\ 
\dot \phi &=\frac{\partial H_{\rm EOB}}{\partial P_\phi}, \dot P_\phi=-\frac{\partial H_{\rm EOB}}{\partial \phi} = 0  
\end{aligned} 
%\right.
\label{eq:EOBdynamics}
\ee
The effective Hamiltonian associated with the deformed-Kerr metric has the form
\be
\hat{H}_{\rm eff}=\frac{\sqrt{\Delta _t \left(\Lambda _t \left(r^2+\Delta _r \hat{P}_r^2\right)+r^4 \hat{P}_{\phi }^2\right)}+\widetilde{\omega}_{\rm fd} \hat{P}_{\phi }}{\Lambda _t},
\label{eq:Heff}
\ee
Solving Eq.(\ref{eq:Heff}) for $P_r$ in terms of $(E,P_\phi,r)$ leads to
\be
\label{eq:prof}\hat{P}_r^2=\frac{\left(\omega _{\text{fd}}^2-r^4 \Delta _t \right)\hat{P}_{\phi }^2+\Lambda _t^2 \hat{H}_{\rm eff}^2-2 \Lambda _t \omega _{\text{fd}} \hat{P}_{\phi }\hat{H}_{\rm eff} -r^2\Delta _t\Lambda _t}{\Delta _r \Delta _t \Lambda _t},
\ee
where we have defined the reduced momenta $\hat{P}_r=P_r/\mu$ and $\hat{P}_\phi=P_\phi/\mu$

The conservative EOB equations of equatorial motion without radiation reaction can then be written as
\bes\label{eomsEOBrphiEL}\bea
\dot r&=&\frac{\nu \Delta _t \Delta _r P_r}{E \sqrt{\Delta _t \left(\Lambda _t \left(r^2+\Delta _r P_r^2\right)+r^4 P_{\phi }^2\right)}} ,  \label{eq:rdoteom}\\
\dot \phi&=&\frac{\nu  \left(\widetilde{\omega}_{\rm fd}+\frac{r^4 \Delta _t P_{\phi } }{\sqrt{\Delta _t \left(\Lambda _t \left(r^2+\Delta _r P_r^2\right)+r^4 P_{\phi }^2\right)}}\right)}{E \; \Lambda _t},\label{eq:phidotgen}
\eea
\ees

Eqs.~\eqref{eomsEOBrphiEL} are more convenient than the original canonical EOB equations \eqref{eq:EOBdynamics}, because the dependence on $\hat{P}_r$ has been eliminated with the energy $E$ which is a constant in conservative system and changes only due to radiation reaction if GW fluxes are considered. \comment{Note that the equations of motion in the form given in Eqs.~\eqref{eomsEOBrphiEL} differ from the canonical EOB dynamics because the dependence on $\hat{P}_r$ has been eliminated in favor of $E$. This is already a more convenient formulation since unlike $\hat{P}_r$, the energy $E$ changes only due to radiation reaction, and the denominators in $\dot{x}^i$ are simple functions of $E$ instead of involving a complicated dependence on all EOB coordinates through $H_{\rm EOB}(x^i, P_i)$ as is the case in the canonical formulation.}

%
%%%%%%%%
\subsection{Re-parameterization of the constants and equation of motion}
\label{sec:constants}
%%%%%%%%
\comment{The description of the dynamics in Eqs.~\eqref{eomsEOBrphiEL} can be further adapted to reflect the properties of the motion in the following way. For an eccentric bound orbit we define the Keplerian orbital elements $p$ and $e$ by} The constants of motion and the dynamical equations in the last subsection can be written in the geometrized orbital elements semilatus rectum $p$ and eccentricity $e$. This will make the description of system more intuitive. For an eccentric orbit, it exists periastron and apastron points which can be expressed as
\be
r_{1}=\frac{p}{1-e}, \ \ \ \ \ r_2=\frac{p}{1+e}, \label{eq:r12}
\ee
where $r_{1,2}$ are the turning points of the radial motion, i.e., apastron and periastron respectively. By setting the radial equation of motion~\eqref{eq:rdoteom} equals to zero with $\hat{P}_r=0$, we can solve out the two points. Furthermore, taking $r_1,~r_2$ into Eq.~\eqref{eq:prof} to make $\hat{P}_r=0$ again, finally we get the constants of motion $(E,\hat{P}_\phi)$ in terms of $(p,e)$
\begin{widetext}
\bes
\label{eq:psofep}\bea
\hat{P}_\phi^2\!&=&\!\frac{\left(a_1\!-\!a_2\right)^2\left(b_1^2\!+\!b_2^2\right)\!-\!(b_1^2\!-\!b_2^2)(b_1^2c_1\!-\!b_2^2c_2)\!-\!2 (a_1\!-\!a_2)b_1 b_2\sqrt{\left(a_1\!-\!a_2\right)^2\!-\!\left(b_1^2\!-\!b_2^2\right)\left(c_1\!-\!c_2\right)}}{\left((a_1\!-\!a_2)^2\!-\!\left(b_1^2 c_1\!-\!b_2^2 c_2\right)\right)^2}  \\
\frac{ E^2}{M^2}\!&=&\!1+2\nu\left(a_1 \Hat{P_{\phi }}\!+\!\sqrt{\frac{c_1 \Hat{P_{\phi }}^2\!+\!1}{b_1} } -1\right)\qquad
\eea
\ees
where the coefficients are
\bes
\bea
a_1&=&\frac{a (1-e)^3 \left((1-e)^2 \nu  \left(\omega_1^{\rm fd}+a^2\omega_2^{\rm fd}\right)+2 p^2\right)}{p^5-a^2 p^3(A(r_1)-2) (1-e)^2 },\\
a_2&=&\frac{a (1+e)^3 \left((1+e)^2 \nu  \left(\omega_1^{\rm fd}+a^2\omega_2^{\rm fd}\right)+2 p^2\right)}{p^5-a^2 p^3(A(r_2)-2) (1+e)^2 },\\
b_1&=&\sqrt{\frac{a^2 (1-e)^2+A(r_1) p^2}{p^2-a^2 (A(r_1)-2) (1-e)^2}},\\
b_2&=&\sqrt{\frac{a^2 (1+e)^2+A(r_2) p^2}{p^2-a^2 (A(r_2)-2) (1+e)^2}},\\
c_1&=&\frac{(1-e)^2}{p^2-a^2 (A(r_1)-2) (1-e)^2},\\
c_2&=&\frac{(1+e)^2}{p^2-a^2 (A(r_2)-2) (1+e)^2}\,.
\eea
\ees
\end{widetext}
The above formalism for Kerr black hole are much more complicated than the Schwarzschild ones in \cite{hinderer2017foundations}. Obviously, for the test-particle limit $\nu \to 0$, the above results will go back the geodesic motion of test particle in Kerr spacetime.

The orbital radius at arbitrary moment is expressed by the semilatus rectum $p$ and the eccentricity $e$ together with the phase variable associated with the spatial geometry of the radial motion denoted by $\xi$. These variables are defined by expressing the radial motion as 
\be
r=\frac{p}{1+e\cos\xi}\,.\\ \label{eq:rofxi}
\ee
so that the periastron and apastron correspond to $\xi=(0, \pi)\,{\rm mod \, 2\pi} $ respectively. Taking derivation on Eq.~(\ref{eq:rofxi}) we will get the evolution equation about the phase variable $\xi$
\be
\label{eq:xidotgen1}
\dot \xi=\frac{(1+e\cos\xi)^2}{epM\sin\xi}\dot r+\frac{\cot\xi}{e}\dot e-\frac{1+e\cos\xi}{ep\sin\xi}\dot p.
\ee
For conservative system $\dot{p}=\dot{e}=0$. If we take the radiation reaction of GWs into account, by differentiating Eqs. (\ref{eq:psofep}), we express the evolution of $(p,~e)$ by the energy and angular momentum fluxes of gravitational radiation
\bes
\bea
\label{eq:edotpdot}
\dot{e}&=&\frac{(\partial E/\partial p)(\dot{P_\phi}/\mu)-(\partial \hat{P}_\phi/\partial p)\dot{E}}{(\partial E/\partial p)(\partial \hat{P}_\phi/\partial e)-(\partial E/\partial e)(\partial \hat{P}_\phi/\partial p)},\\
\dot{p}&=&\frac{(\partial \hat{P}_\phi/\partial e)\dot{E}-(\partial E/\partial e)(\dot{P_\phi}/\mu)}{(\partial E/\partial p)(\partial \hat{P}_\phi/\partial e)-(\partial E/\partial e)(\partial \hat{P}_\phi/\partial p)}\,.
\eea
\ees

The final set of EOB equations of motion with radiation reaction are Eqs.~\eqref{eq:edotpdot} together with the evolution of the phases described by Eq. \ref{eq:xidotgen1} and Eq. \ref{eq:phidotgen}, and the radius of motion at arbitrary moment is given by Eq. \ref{eq:rofxi}. Now all the equations of motion are expressed in terms of only the geometric parameters $(p,e,\xi)$ and effective Kerr parameter $a$.  

Finally we can get the orbital coordinates of the effective test particle in terms of only $\xi$: 
\bes
\bea
\label{def_orbitaleq_in}
t(\xi) &=&\int_0^{\xi}\frac{1}{\cal P}d\xi\,,\\
r(\xi) &=&\frac{p }{1+e\cos\xi} \,,\\
\theta &=& \frac{\pi}{2}\,,\\
\phi(\xi)&=&\int_0^{\xi}\frac{\dot\phi}{\cal P}d\xi\,,\label{orbital equations}
\eea
where ${\cal P}(e,p,\xi)$ is the first term in the right hand of Eq. (\ref{eq:xidotgen1})
\be
  {\cal P}(e,p,\xi)=\frac{(1+e\cos\xi)^2}{ep\sin\xi} \times \frac{\nu \Delta _t \Delta _r P_r}{E \sqrt{\Delta _t \left(\Lambda _t \left(r^2+\Delta _r P_r^2\right)+r^4 P_{\phi }^2\right)}}\,. 
\ee
\ees

%%%%%%%%
 \subsection{Quantitative influences of mass-ratio on the conservative dynamics}
 \label{sec:frequencies}
%%%%%%%%
%Eccentric planar orbits possess two frequencies characterizing the radial librations between the turning points and the azimuthal rotations. In the Newtonian limit both of these frequencies coincide, however, this degeneracy is broken for relativistic motion. The frequencies are defined as follows. One period of the radial motion is the time elapsed between successive periapsis passages, and hence the time taken for $\xi$ to increase from $0$ to $2\pi$. From the conservative part of Eq.~(\ref{eq:dotxi}), the corresponding radial frequency is given by
 %
 Now we calculate the fundamental orbital frequencies in terms of the geometric parameters. This was done in the test particle limit due to the analytical integrals of geodesic in Kerr spacetime. While in the EOB formalism with the mass-ratio correction, the situation becomes complicated. Firstly, we express the radial frequency $\omega_r$ which reflect the period of the radial motion from the periastron to apastron and back to periastron again, and the orbital period $T_r$ can be calculated by taking $\xi$ from $0$ to $2\pi$, then
 \be
 \omega_r=\frac{2\pi}{\int^{2\pi}_0 \frac{1}{\cal P}d\xi}=\frac{2 \pi }{\int_0^{2 \pi } \frac{ep\sin\xi}{(e\cos\xi+1)^2\dot{r}} \, d\xi }. \label{eq:omegardef}
 \ee
In Eq. (\ref{eq:phidotgen}), we have already written the variation of $\phi$ with coordinate time $t$. Rigidly, the radial motion is the real periodic motion. When the particle passes through the periastron twice, $\Delta \phi$ will be larger than $2 \pi$ because of the periastron procession caused by relativistic effect. We can then define the frequency of the azimuthal motion by $\Delta \phi/T_r$. So, we compute the azimuthal frequency from the orbit-average of the $\phi$ motion as
 \be
   \omega_\phi=\langle \dot{\phi}\rangle=\frac{\int^{2\pi}_0\frac{\dot{\phi}}{\cal P}d\xi}{\int^{2\pi}_0 \frac{1}{\cal P}d\xi}=\frac{\int_0^{2 \pi } \frac{\dot{\phi} e p\sin\xi}{(e\cos\xi+1)^2\dot{r}} \, d\xi }{\int_0^{2 \pi} \frac{ep\sin\xi}{(e\cos\xi+1)^2\dot{r}}\, d\xi }. \label{eq:omegaphidef}
   \ee
Where $\dot{r}$ is given in Eq. (\ref{eq:rdoteom}), and $P_r, ~E,~P_\phi$ are expressed in Eqs. (\ref{eq:prof},\ref{eq:psofep}), just replace $r$ in these equation with $\xi$, the above two integrals only contain argument $\xi$ and can be integrated easily. Here we do not write down the fully expanded expressions of these two integrals, it is direct and trivial.
%\subsection{Illustrative results}

\begin{figure}
\includegraphics[width=0.9\columnwidth]{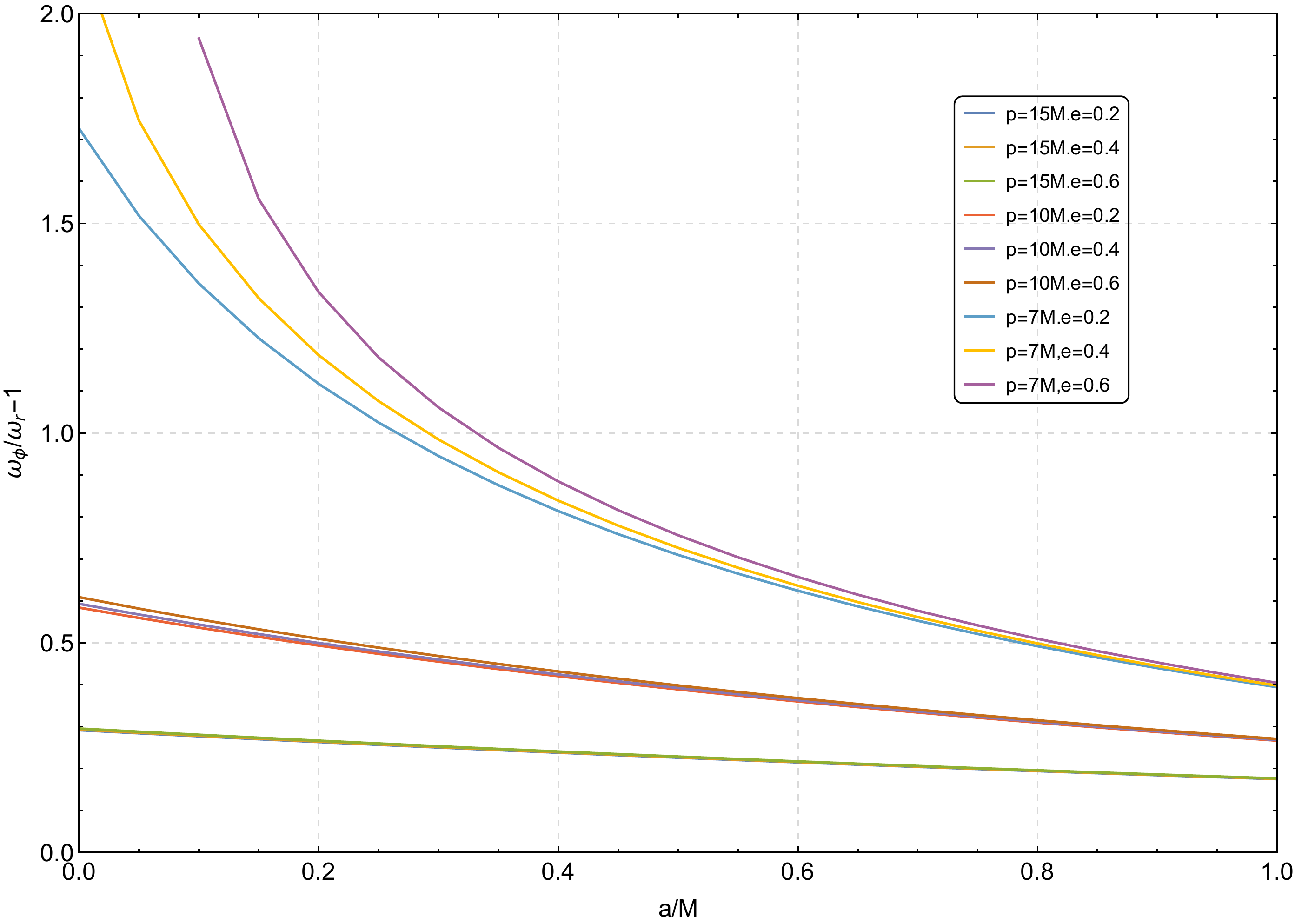}
\caption{\emph{Effect of spin parameter on periastron procession for an binary with symmetric mass-ratio $\nu=10^{-3}$}. At a fixed spin parameter, an orbit with bigger semilatu rectums has a smaller periastron procession than those with smaller semilatus rectums. For the orbits with the same eccentricity and semilatus rectum, the higher spin parameter induces a smaller periastron precession.}
\label{fig:1}
\end{figure}

With the expressions of two fundamental frequencies at hand, now we investigate how the effective Kerr parameter $a$ and symmetric mass-ratio $\nu$ affect on the features of the radial and azimuthal frequencies from the test particle limit in conservative dynamics. The effect of spin parameter on these frequencies for various mean orbital separation and eccentricity is shown in Fig.~\ref{fig:1}. Obviously, as the semilatus rectum increases,in other words, mean orbital radius correspondingly increases, this leads to a decrease of the periastron procession $\omega_\phi/\omega_r-1$, approaches to the Newtonian limit $\omega_r=\omega_\phi$. Interestingly, as the spin of central SMBH increases, the periastron procession decreases. The effect of eccentricity on the procession becomes obvious only when the separation $p$ and spin $a$ are both small (see the top three curves in Fig.~\ref{fig:1}). In this extreme mass-ratio case ($\nu = 10^{-3}$), the behaviour of periastron procession depending on the spin, orbital separation and eccentricity are similar. 

Next, we consider the radial and azimuthal frequencies' shifts due to the mass-ratio of the binary in the absence of radiation reaction. As we all known, the test-particle orbit already has a precise analytical model. For the binary system which consider mass radio, now we have the analytical EOB orbital solution with eccentricity for spinning BHs. In order to observe the impact of mass-ratio on radial and azimuthal frequencies in different conditions with various orbital parameters, we compare the test-particle results and the EOB results. This comparison is shown in Table~\ref{frequency comp}. The right-most column of this table is the relative frequency shift divided by mass-ratio $\Delta\omega/(\omega\nu)$, where $\Delta \omega/\omega$ is the relative difference of radial/azimuthal frequency between the test-particle frequency $(\omega _{r0}, \omega _{\phi0})$ and the EMRI ones, i.e., $(\omega _{r}-\omega _{r0})/\omega _{r0}, (\omega _{\phi}-\omega _{\phi0})/\omega _{\phi0}$.

\begin{table*}
\caption{The relative differences of orbital frequencies $\omega _r,~\omega _\phi$ between test-particle and EOB models with small mass-ratios.}
\begin{tabular}{c|c|c|c|c|c|c|c|c|c|c|c|c}
\hline$a/M$&$p/M$&$e$&&test-particle&$\nu=10^{-2}$&$\nu=10^{-3}$&$\nu=10^{-4}$&$\nu=10^{-5}$&$\nu=10^{-6}$&$\frac{\Delta \omega}{\omega}(/\nu)$\\ 
\hline
\multirow{2}{*} {$0.99$} &\multirow{2}{*} {$5$}&\multirow{2}{*} {$0.1$}&$\omega _r$&$0.050841032$&$0.052966704$&$0.051054728$&$0.050862405$&$0.050843169$&$0.050841245$&$4.2$\\ 
 &&&$\omega _\phi$&$0.0081375480$&$0.0080796035$&$0.008.1318547$&$0.0081369800$&$0.0081374912$&$0.0081375423$&$0.8$\\  \hline
\multirow{2}{*} {$0.99$} &\multirow{2}{*}{$10$}&\multirow{2}{*}{$0.1$} &$\omega _r$&$0.023863900$&$0.023996267$&$0.023876962$&$0.023865204$&$0.023864030$&$0.023863913$&$0.55$ \\
&&& $\omega _\phi$&$0.030284712$&$0.030221400$&$0.030278593$&$0.030284102$&$0.030284651$&$0.030284706$&$0.20$\\ \hline
\multirow{2}{*} {$0.99$} &\multirow{2}{*} {$5$}&\multirow{2}{*}{$0.6$} & $\omega _r$&$0.031648541$&$0.032570745$&$0.031742354$&$0.031657935$&$0.031649480$&$0.031648635$&$3.0$ \\ 
 &&&$\omega _\phi$&$0.051669258$&$0.050273106$&$0.051525327$&$0.051654825$&$0.051667814$&$0.051669114$&$2.8$\\ \hline
\multirow{2}{*} {$0.99$} &\multirow{2}{*} {$20$}&\multirow{2}{*} {$0.6$}&$\omega _r$&$0.0052501837$&$0.0052537835$&$0.0052505381$&$0.0052502191$&$0.0052501873$&$0.0052501841$&$0.13$\\ 
 &&&$\omega _\phi$&$0.0059449013$&$0.0059371680$&$0.0059441520$&$0.0059448266$&$0.0059448938$&$0.0059449005$&$0.067$\\ \hline
\multirow{2}{*} {$0.5$} &\multirow{2}{*}{$5$}&\multirow{2}{*} {$0.1$}& $\omega _r$&$0.030270602$&$0.032859193$&$0.030538216$&$0.030297442$&$0.030273287$&$0.030270870$&$8.9$ \\ 
&&& $\omega _\phi$&$0.085402159$&$0.084958246$&$0.085355748$&$0.085397497$&$0.085401693$&$0.085402113$&$0.55$\\ \hline
\end{tabular}
\label{frequency comp}
\end{table*}

The frequency shift $\Delta\omega/(\omega\nu)$ due to the mass-ratio is almost independent with mass-ratio itself, and the relative shift $\Delta\omega/\omega$ is in a range about $[0.1\nu,~10\nu]$ based on Table \ref{frequency comp}. This indicates that we must consider the influence of mass-ratio in the conservative dynamics for EMRIs, because the cycles of an typical EMRI waves $\sim 1/\nu$ in LISA band, and if the relative error of frequency reaches $\sim \nu$, the dephase will accumulate to a few of radians at the end of evolution. This may induce a failure of detection of EMRIs with the test particle model.

\begin{figure}
\centering  
\subfigure{\includegraphics[width=0.45\columnwidth]{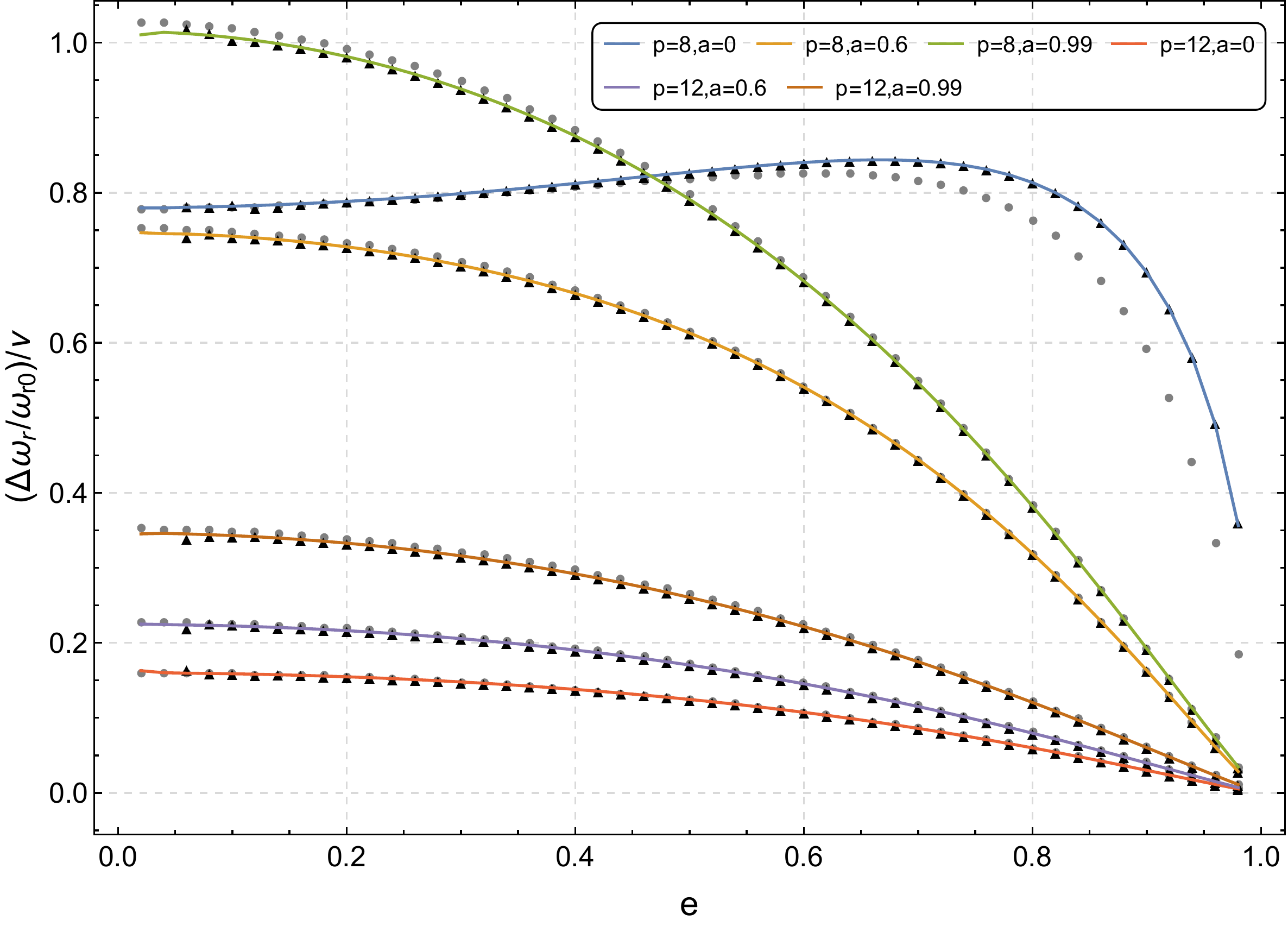}}
\subfigure{\includegraphics[width=0.45\columnwidth]{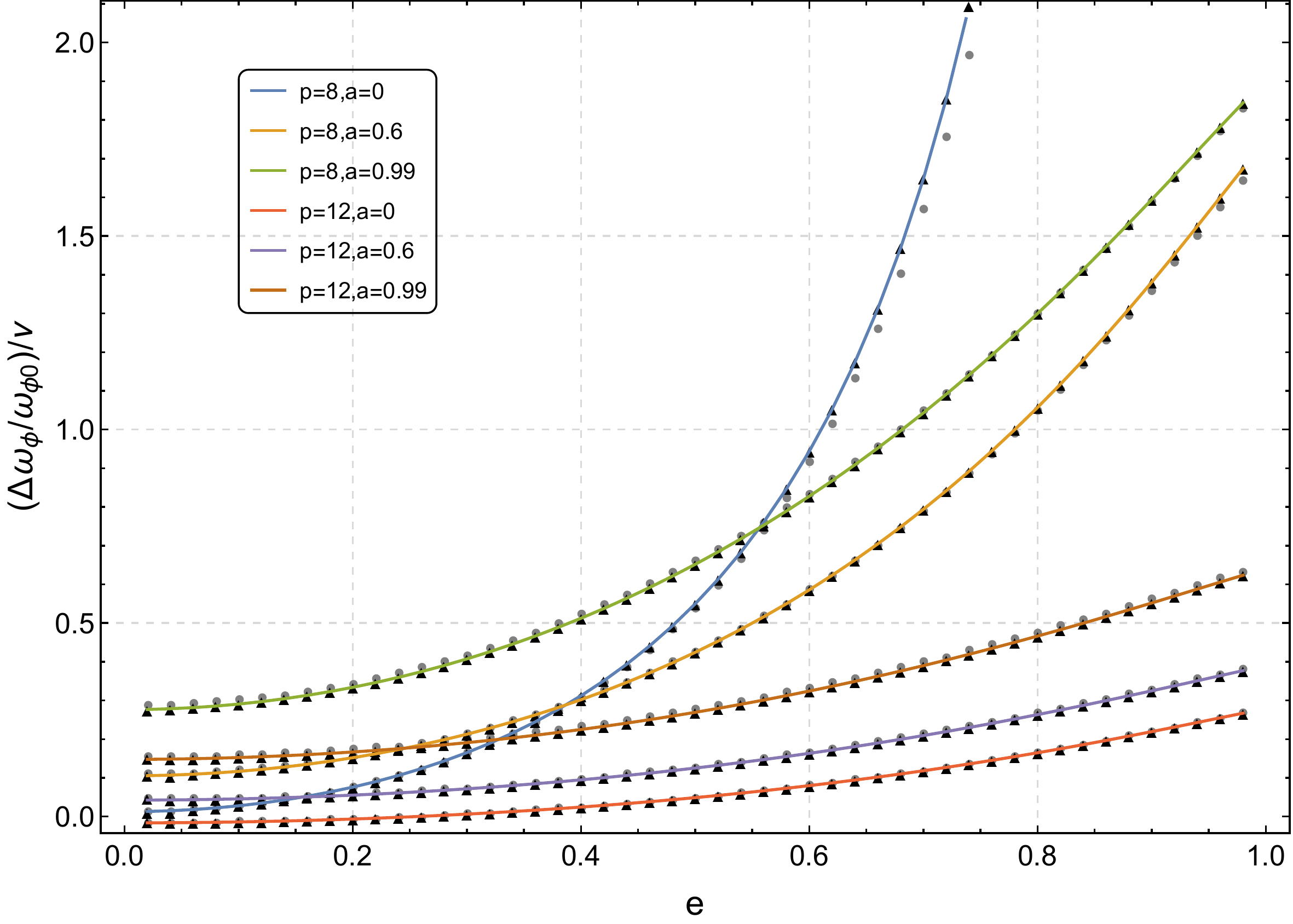}}
\caption{\emph{The frequency shifts $\frac{\Delta\omega_{r}}{\nu\omega_{r0}}$ and $\frac{\Delta\omega_{\phi}}{\nu\omega_{\phi0}}$ versus eccentricity $e$ in the cases of various $a,~\nu,~ p$.} The triangles, solid line and points represent $\nu = 10^{-6}, 10^{-4}$ and $10^{-2}$ respectively.}
\label{fig:2}
\end{figure}

Figure~\ref{fig:2} illustrates the effect of eccentricity on the radial frequency shift(left panel) and the azimuthal one (right panel) with various semilatus rectum $p$ and spin parameter $a$. It shows that the shifts of $\omega_r$ and $\omega_\phi$ due to mass-ratio have different performances versus eccentricity. The results for different mass-ratios are also plotted (triangles, solid line and points represent $\nu = 10^{-6}, 10^{-4}$ and $10^{-2}$ respectively), clearly show that $\Delta \omega/(\omega \nu)$ is not sensitive to mass-ratio except for the radial frequency while the trajectory approaches the vicinity of the innermost stable orbit (ISO) and mass-ratio becomes 0.01 (blue lines in the left panel). 

\begin{figure}
\centering  
\subfigure{\includegraphics[width=0.45\columnwidth]{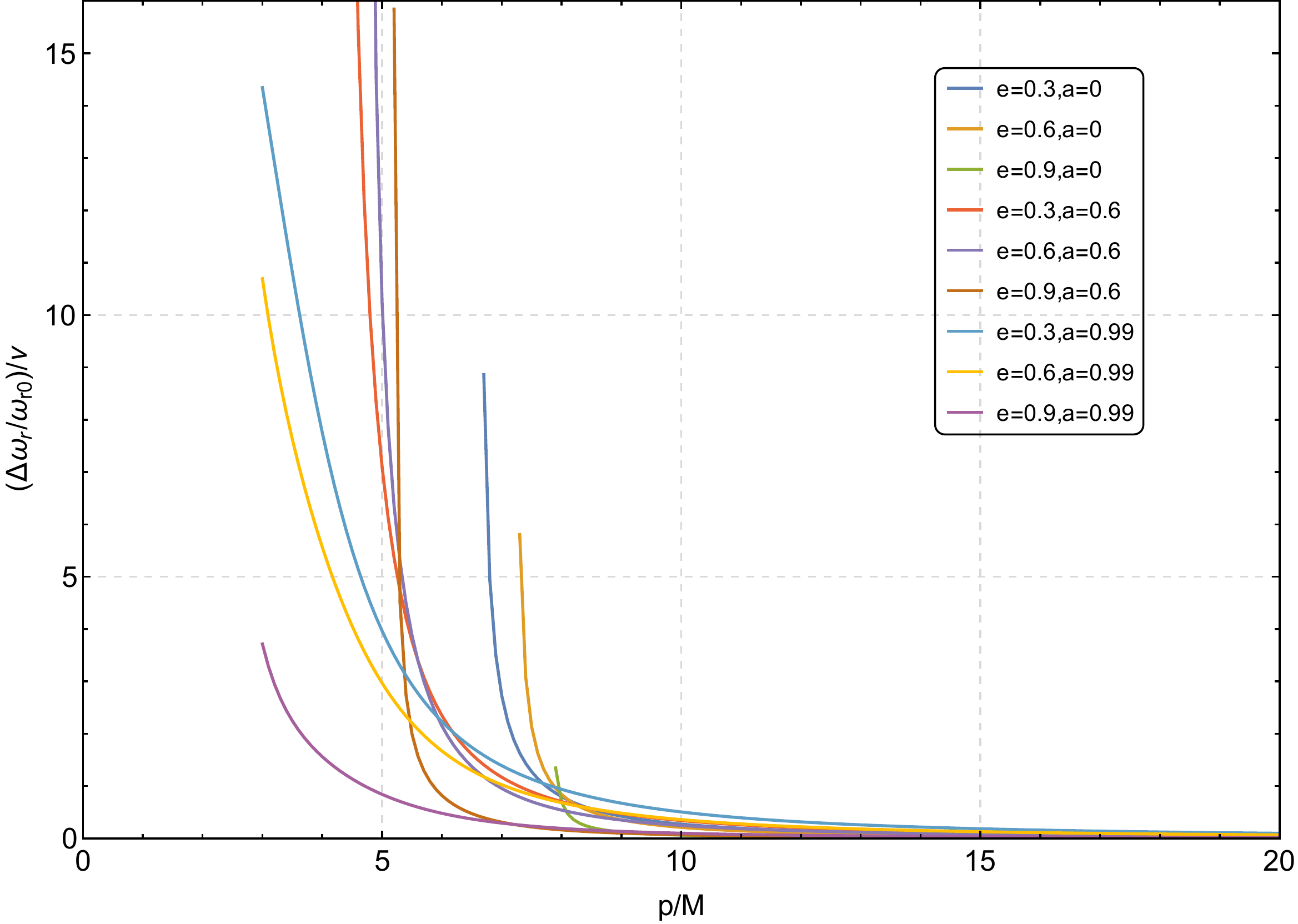}}
\subfigure{\includegraphics[width=0.45\columnwidth]{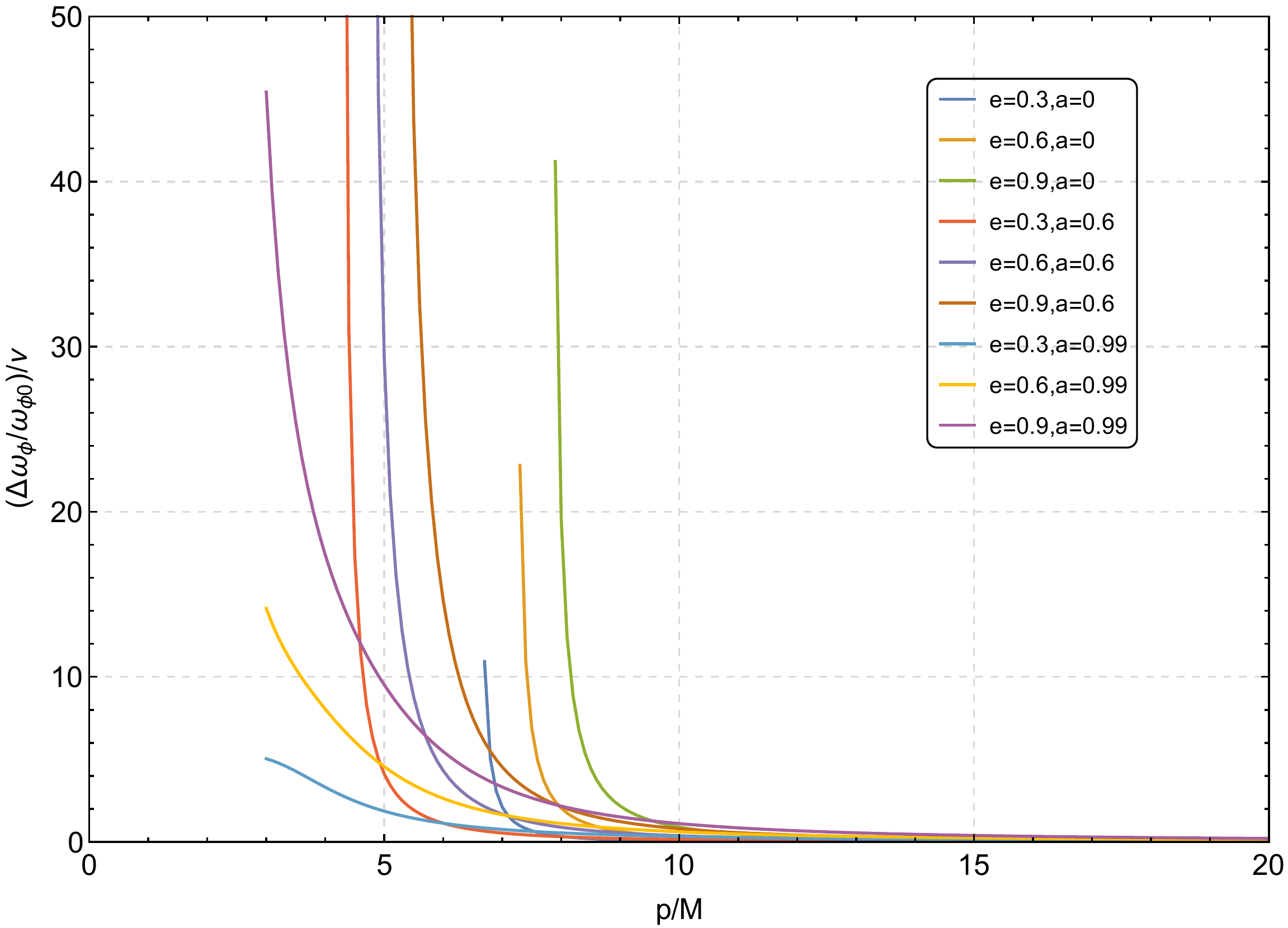}}
\caption{\emph{The frequency shifts $\frac{\Delta\omega_{r}}{\nu\omega_{r0}}$ and $\frac{\Delta\omega_{\phi}}{\nu\omega_{\phi0}}$ versus semilatus rectum $p$ in the cases of various $a,~ e$ with $\nu = 10^{-4}$.} When $p$ becomes small, the frequency shift grows very fast.}
\label{fig:3}
\end{figure}

Figure~\ref{fig:3} demonstrates the effect of semilatus rectum $p$ on the radial frequency shift(left panel) and the azimuthal one (right panel) with various spin $a$ and eccentricity $e$. When $p$ becomes smaller, the frequency shifts become larger. The sudden growth of frequency shift is due to the orbit approaching to the ISO.

\begin{figure}
\centering  
\subfigure{\includegraphics[width=0.45\columnwidth]{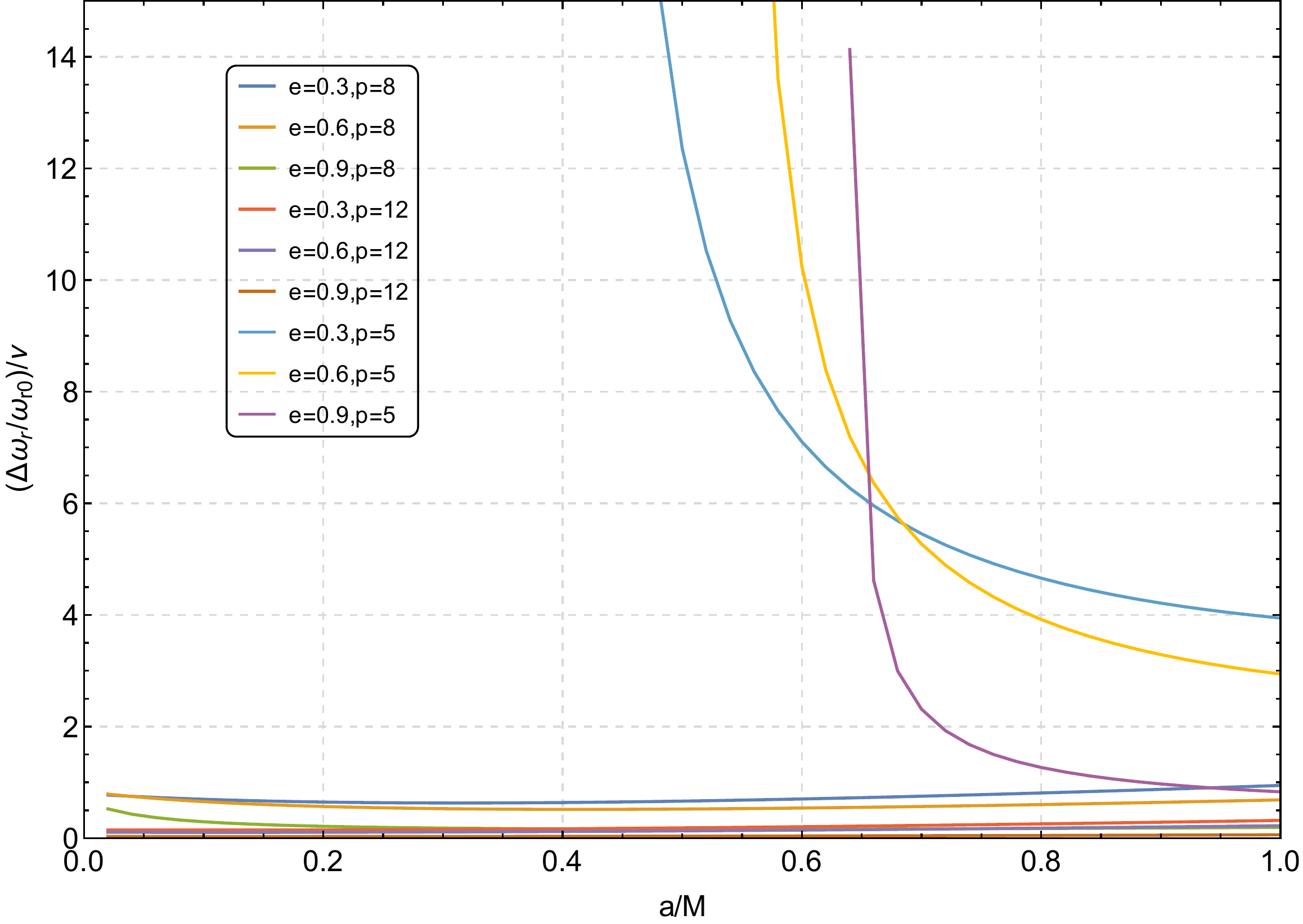}}
\subfigure{\includegraphics[width=0.45\columnwidth]{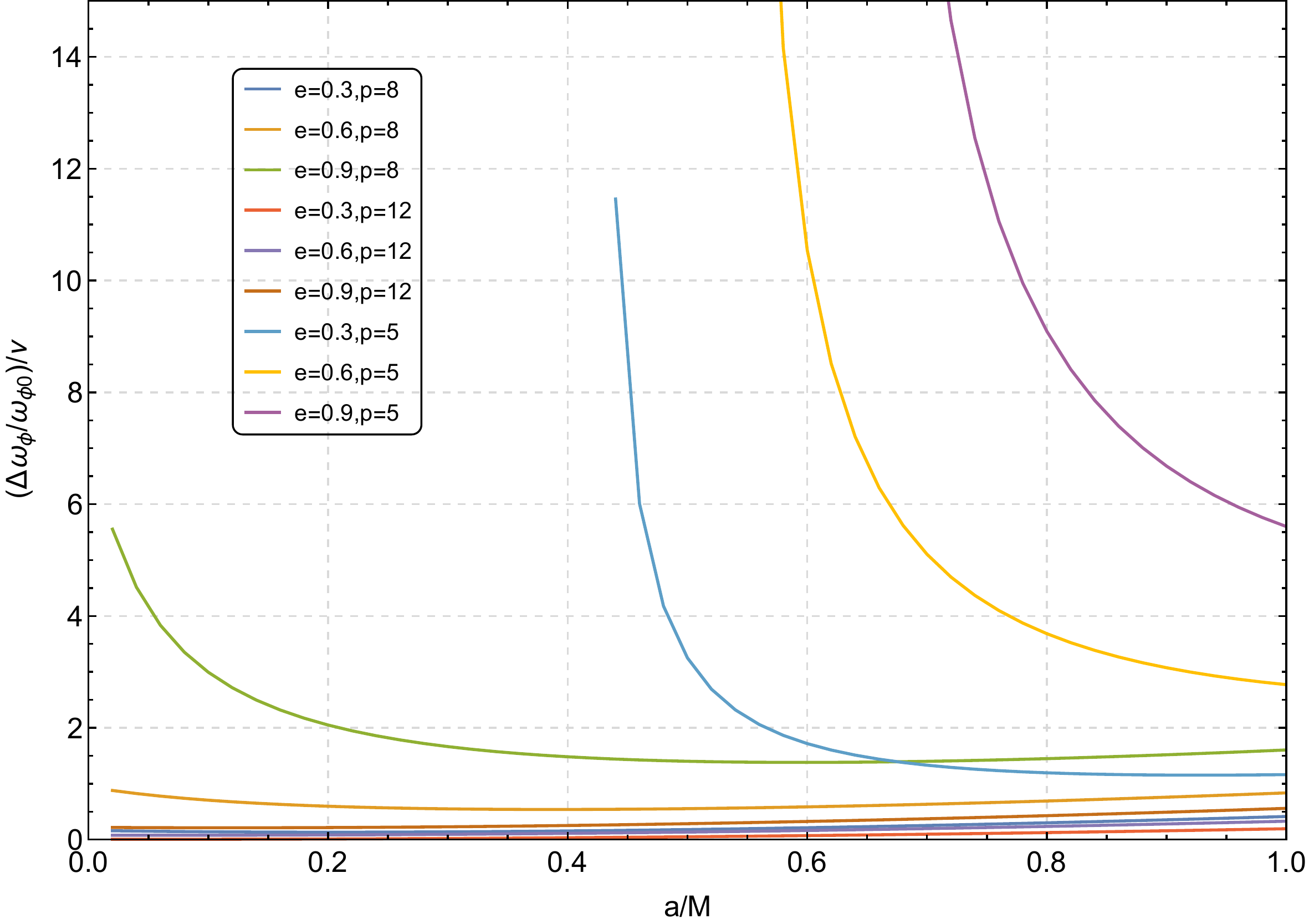}}
\caption{\emph{The frequency shifts $\frac{\Delta\omega_{r}}{\nu\omega_{r0}}$ and $\frac{\Delta\omega_{\phi}}{\nu\omega_{\phi0}}$ versus Kerr parameter $a$ in the cases of various $p,~ e$ with $\nu = 10^{-4}$.} When $a$ becomes smaller, the frequency shift could become larger due to the orbit approaching ISO.}
\label{fig:4}
\end{figure}

Figure~\ref{fig:4} illustrates the effect of BH's spin on the radial frequency shift (left panel) and the azimuthal one (right panel) with various semilatus rectum $p$ and eccentricity $e$. For the cases of $p = 5$, when $a$ becomes small, the orbits will be very close to the ISO, then the frequency shifts grow very fast.

\begin{figure}
\includegraphics[width=0.9\columnwidth]{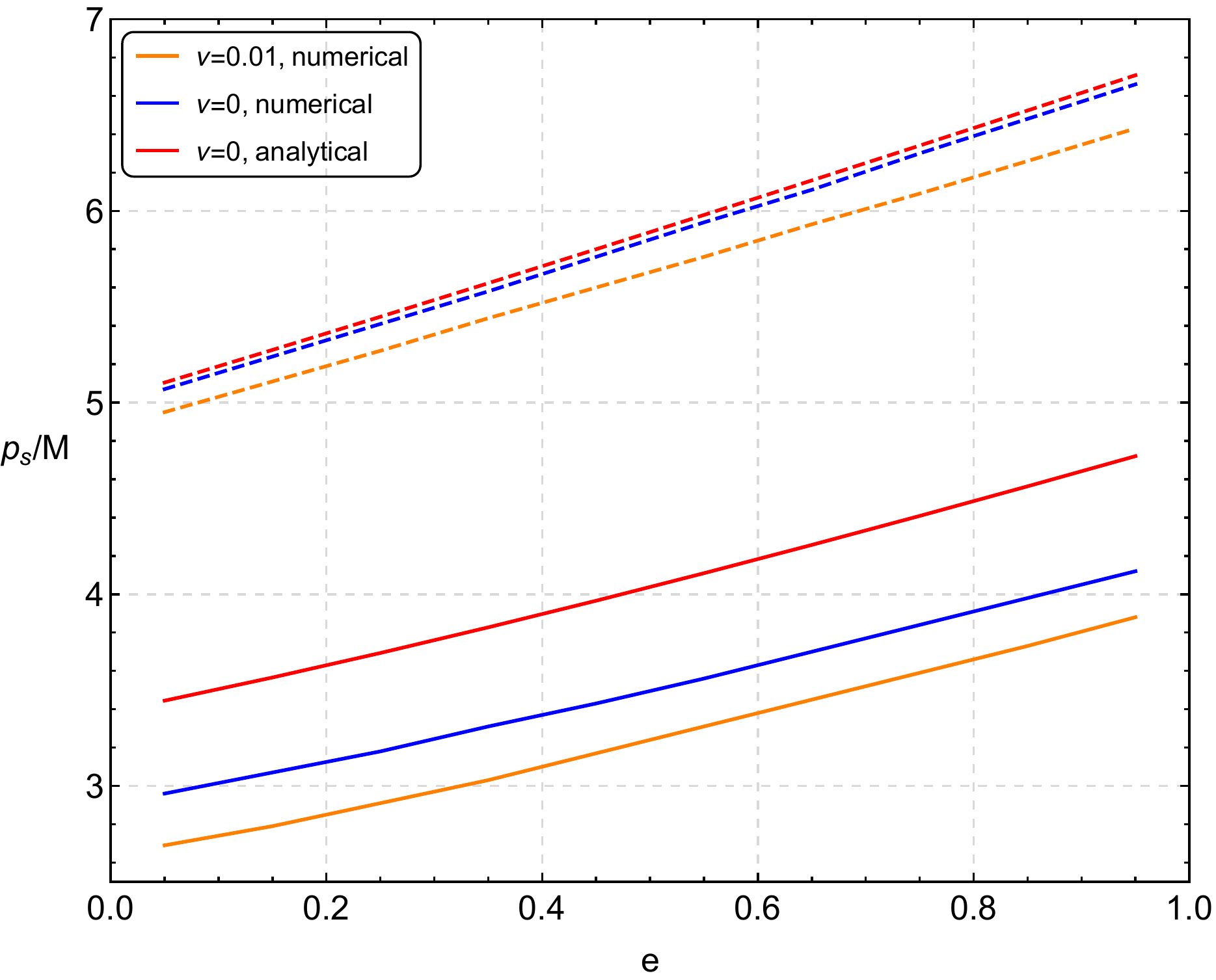}
\caption{\emph{The effect of mass-ratio on the boundary curve $p_s(e)$}. The solid and dashed lines represent the case of $a=0.8$ and $a=0.3$ respectively. The analytical results are obtained from Eq.(\ref{ps}), and the numerical ones are calculated from the equations of motion.}
\label{fig:5}
\end{figure}

The semilatus rectum $p$ of ISO is the separatrix of bound orbits and in test particle limit is given by the analytic expression from Eq.(24) of~\cite{glampedakis2002zoom}  
\be
p_s=(6+2a) M\mp 8 a \sqrt{\frac{1+e}{2 e+6}}+\mathcal{O}(a^2) \,. \label{ps}
\ee
However, the above equation is approximate of ISO even for test particles. Considering the influence of the small mass on the background, the ISO of EMRIs should deviate from the test particle one.  
Figure~\ref{fig:5} shows the effect of mass-ratio on the boundary curve $p_s(e)$. We can see that the deviation of EOB's ISO can be 10\% of the test particle model. In addition, the error of approximate expression (\ref{ps}) becomes large for fast spinning BH.

%%%%%%%%%%%%%%%%%%%%%%%%%%%%%%%%%%%%%%%%%%%%%%%%%%%%%%
\section{The orbital evolution and waveforms}
\label{sec:fluxes}

In this section we introduce the gravitational wave fluxes of energy and angular momentum in previous literature, and calculate the gravitational wave strain by the Teukolsky equation which is a perturbation theory of Kerr black hole\cite{Teukolsky}. We include the mass-ratio into the 1PN terms of energy and angular momentum fluxes, and compare the performance of approximate fluxes from 2PN to 4PN. 
%%%%%%%%%%%%%%%%%%%%%%%%%%%%%%%%%%%%%%%%%%%%%%%%%%%%%%
\subsection{Radiation fluxes}
\label{sec:instantfluxes}
%%%%%%%%%%%%%%%%%%%%%%
The analytic 4PN $\mathcal{O}(e^6)$ formulae of energy and angular momentum fluxes in Boyer-Lindquist coordinates are given by~\cite{sago2015calculation} in parameter $\upsilon \equiv \sqrt{1/p}$. For convenience, we use $p$ and transfer their expressions as follows  

\begin{widetext}

\bea 
\langle{\cal F}\rangle_{3{\rm PN}}&=&\frac{32\mu^2(1-e^2)^{3/2}}{5M^2 p^5}\bigg\{1+\frac{73}{24}e^2+\frac{37}{96}e^4-\frac{1}{p}\bigg[\frac{1247}{336}+\frac{9181e^2}{672}-\frac{809e^4}{128}-\frac{8609e^6}{5376}\bigg]\nonumber\hspace{0.5cm}\\
&&+\frac{1}{p^{3/2}}\bigg[\pi  \left(4+\frac{1375 e^2}{48}+\frac{3935 e^4}{192}+\frac{10007 e^6}{9216}\right)-q \left(\frac{73}{12}+\frac{823 e^2}{24}+\frac{949 e^4}{32}+\frac{491 e^6}{192}\right)\bigg] \nonumber\hspace{0.5cm}\\
&&-\frac{1}{p^2}\bigg[\frac{44711}{9072}+\frac{172157 e^2}{2592}+\frac{2764345 e^4}{24192}-\frac{3743 e^6}{2304}-q^2\left(\frac{33}{16}+\frac{359 e^2}{32}+\frac{1465 e^4}{128}+\frac{883 e^6}{768}\right)\bigg]\nonumber\hspace{0.5cm}\\
&&-\frac{1}{p^{5/2}}\bigg[\pi \left(\frac{8191}{672}+\frac{44531 e^2}{336}+\frac{4311389 e^4}{43008}-\frac{15670391 e^6}{387072}\right)\nonumber\hspace{0.5cm}\\ 
&&-q\left(\frac{3749}{336}+\frac{1759e^2}{56}-\frac{111203 e^4}{1344}-\frac{49685e^6}{448}\right)\bigg]\nonumber\hspace{0.5cm}\\
&&+\frac{1}{p^3}\bigg[\frac {6643739519}{69854400}+\frac {43072561991e^2}{27941760}+\frac {919773569303e^4}{279417600}+\frac {308822406727e^6}{186278400}\nonumber\hspace{0.5cm}\\
&&-\gamma\left(\frac {1712}{105}+\frac {14552e^2}{63}+\frac {553297e^4}{1260}+\frac {187357e^6}{1260}\right)\nonumber\hspace{0.5cm}\\ 
&&-\ln\left(2\right)\left(\frac{3424}{105}+\frac {13696e^2}{315}+\frac {12295049e^4}{1260}-\frac{24908851e^6}{252}\right)\nonumber\hspace{0.5cm}\\
&&-\ln\left(3\right)\left(\frac {234009e^2}{560}-\frac {2106081e^4}{448}+\frac {864819261e^6}{35840}\right)-\frac {5224609375e^6}{193536}\ln\left(5\right) \nonumber \hspace{0.5cm}\\
&&+\pi^{2}\left( \frac{16}{3}+\frac {680e^2}{9}+\frac {5171e^4}{36}+\frac {1751e^6}{36}\right)-q\pi\left(\frac {169}{6}+\frac {4339e^2}{16}+\frac {42271e^4}{96}+\frac {4867907e^6}{27648}\right)\nonumber \hspace{0.5cm}\\
&&+q^2\left(\frac{3419}{168}+\frac{50271e^2}{224}+\frac{340141e^4}{896}+\frac{1013347e^6}{5376}\right) \nonumber \hspace{0.5cm}\\
&&+ \ln p\left(\frac {856}{105}+\frac {7276e^2}{63}+\frac {553297e^4}{2520}+\frac {187357e^6}{2520}\right)\bigg]\bigg\},\\
\langle{\cal G}^z\rangle_{3{\rm PN}}&=&\frac{32\mu^2(1-e^2)^{3/2}}{5M^2 p^{7/2}}\bigg\{1+\frac{7}{8}e^2-\frac{1}{p}\bigg[\frac{1247}{336}+\frac{425e^2}{336}-\frac{10751e^4}{2688}\bigg]\nonumber\hspace{0.5cm}\\
&&+\frac{1}{p^{3/2}}\bigg[\pi  \left(4+\frac{97 e^2}{8}+\frac{49 e^4}{32}-\frac{49 e^6}{4608}\right)-q \left(\frac{61}{12}+\frac{119 e^2}{8}+\frac{183 e^4}{32}\right)\bigg]\nonumber\hspace{0.5cm}\\
&&-\frac{1}{p^2}\bigg[\frac{44711}{9072}+\frac{302893 e^2}{6048}+\frac{701675 e^4}{24192}-\frac{162661 e^6}{16128}-q^2\left(\frac{33}{16}+\frac{95 e^2}{16}+\frac{311 e^4}{128}\right)\bigg]\nonumber\hspace{0.5cm}\\
&&-\frac{1}{p^{5/2}}\bigg[\pi\left(\frac{8191}{672}+\frac{48361 e^2}{1344}-\frac{1657493 e^4}{43008}-\frac{5458969 e^6}{774144}\right)-q\left(\frac{417}{56}-\frac{5441e^2}{672}-\frac{1097e^4}{24}-\frac{153605e^6}{5376}\right)\bigg]\nonumber\hspace{0.5cm}\\
&&+\frac{1}{p^3}\bigg[\frac {6643739519}{69854400}+\frac {6769212511e^2}{8731800}+\frac {4795392143e^4}{7761600}+\frac {
31707715321e^6}{186278400}\nonumber\hspace{0.5cm}\\
&&-\gamma\left(\frac {1712}{105}+\frac {24503e^2}{210}+\frac {11663e^4}{140}+\frac {2461e^6}{560}\right)\nonumber\hspace{0.5cm}\\ 
&&-\ln\left(2\right)\left(\frac{3424}{105}-\frac {1391e^2}{30}+\frac {418049e^4}{84}-\frac{94138279e^6}{2160}\right)\nonumber\hspace{0.5cm}\\
&&-\ln\left(3\right)\left(\frac {78003e^2}{280}-\frac {3042117e^4}{1120}+\frac {42667641e^6}{3584}\right)-\frac {1044921875e^6}{96768}\ln\left(5\right) \nonumber \hspace{0.5cm}\\
&&+\pi^{2}\left( \frac{16}{3}+\frac {229e^2}{6}+\frac {109e^4}{4}+\frac {23e^6}{16}\right)-q\pi\left(\frac{145}{6}+\frac{409e^2}{3}+\frac{22631e^4}{192}+\frac{69887e^6}{6912}\right)\nonumber \hspace{0.5cm}\\
&&+q^2\left(\frac{799}{56}+\frac{6213e^2}{56}+\frac{143159e^4}{1344}+\frac{14447e^6}{448}\right) \nonumber \hspace{0.5cm}\\
&&+ \ln p\left(\frac {856}{105}+\frac {24503e^2}{420}+\frac {11663e^4}{280}+\frac {2461e^6}{1120}\right)\bigg]\bigg\}\,.
\eea
where $q=a/M$ is the dimensionless spin. The averages of the fluxes at 1PN order are given in terms of the quantities $(e,p)$ and the mass-ratio $\nu$~\cite{hinderer2017foundations}
\bea 
\langle{\cal F}\rangle&=&\frac{32\mu^2(1-e^2)^{3/2}}{5M^2 p^5}\bigg\{1+\frac{73}{24}e^2+\frac{37}{96}e^4\nonumber\hspace{0.5cm}\\
&&-\frac{1}{p}\bigg[\frac{1247}{336}+\frac{5\nu}{4}+e^2\left(\frac{9181}{672}+\frac{325\nu}{24}\right)-e^4\left(\frac{809}{128}-\frac{435\nu}{32}\right)-e^6\left(\frac{8609}{5376}-\frac{185\nu}{192}\right)\bigg]\bigg\},\\
\langle{\cal G}^z\rangle&=&\frac{32\mu^2(1-e^2)^{3/2}}{5M^2 p^{7/2}}\bigg\{1+\frac{7}{8}e^2\nonumber\hspace{0.5cm}\\
&&-\frac{1}{p}\bigg[\frac{1247}{336}+\frac{7\nu }{4}+e^2 \left(\frac{425}{336}+\frac{401\nu}{48}\right)-e^4\left(\frac{10751}{2688}-\frac{205 \nu }{96}\right)\bigg]\bigg\}\,.
\eea
By combining the PN fluxes of a test-particle orbiting a Kerr black hole and the 1PN fluxes of a nonspinning binary with mass-ratio, we derive expressions for energy and angular momentum fluxes contain both spin and mass-ratio
\bea 
\langle{\cal F}\rangle_{2{\rm PN}}&=&\frac{32\mu^2(1-e^2)^{3/2}}{5M^2 p^5}\bigg\{1+\frac{73}{24}e^2+\frac{37}{96}e^4\nonumber\hspace{0.5cm}\\
&&-\frac{1}{p}\bigg[\frac{1247}{336}+\frac{5\nu}{4}+e^2\left(\frac{9181}{672}+\frac{325\nu}{24}\right)-e^4\left(\frac{809}{128}-\frac{435\nu}{32}\right)-e^6\left(\frac{8609}{5376}-\frac{185\nu}{192}\right)\bigg]\nonumber\hspace{0.5cm}\\
&&+\frac{1}{p^{3/2}}\bigg[\pi  \left(4+\frac{1375 e^2}{48}+\frac{3935 e^4}{192}+\frac{10007 e^6}{9216}\right)-q \left(\frac{73}{12}+\frac{823 e^2}{24}+\frac{949 e^4}{32}+\frac{491 e^6}{192}\right)\bigg] \nonumber\hspace{0.5cm}\\
&&-\frac{1}{p^2}\bigg[\frac{44711}{9072}+\frac{172157 e^2}{2592}+\frac{2764345 e^4}{24192}-\frac{3743 e^6}{2304}-q^2\left(\frac{33}{16}+\frac{359 e^2}{32}+\frac{1465 e^4}{128}+\frac{883 e^6}{768}\right)\bigg]\bigg\},\\
\langle{\cal G}^z\rangle_{2{\rm PN}}&=&\frac{32\mu^2(1-e^2)^{3/2}}{5M^2 p^{7/2}}\bigg\{1+\frac{7}{8}e^2\nonumber\hspace{0.5cm}\\
&&-\frac{1}{p}\bigg[\frac{1247}{336}+\frac{7\nu }{4}+e^2 \left(\frac{425}{336}+\frac{401\nu}{48}\right)-e^4\left(\frac{10751}{2688}-\frac{205 \nu }{96}\right)\bigg]\nonumber\hspace{0.5cm}\\
&&+\frac{1}{p^{3/2}}\bigg[\pi  \left(4+\frac{97 e^2}{8}+\frac{49 e^4}{32}-\frac{49 e^6}{4608}\right)-q \left(\frac{61}{12}+\frac{119 e^2}{8}+\frac{183 e^4}{32}\right)\bigg]\nonumber\hspace{0.5cm}\\
&&-\frac{1}{p^2}\bigg[\frac{44711}{9072}+\frac{302893 e^2}{6048}-\frac{701675 e^4}{24192}+\frac{162661 e^6}{16128}-q^2\left(\frac{33}{16}+\frac{95 e^2}{16}+\frac{311 e^4}{128}\right)\bigg]\bigg\}\,.
\eea
\end{widetext}
Here for simplicity, we just write the 2PN formalism. In the adiabatic limit, the evolution of energy and angular momentum is driven by the orbit-averaged radiation reaction forces so that
\be
\dot{E}=-\langle{\cal F}\rangle, \quad \dot{P_\phi}=-\langle{\cal G}^z\rangle,
\ee
Then the evolution of $e$ and $p$ due to the gravitational radiation is given by 
\bes
\bea
\dot{e}&=&\frac{(\partial E/\partial p)(\dot{P_\phi}/\mu)-(\partial \hat{P}_\phi/\partial p)\dot{E}}{(\partial E/\partial p)(\partial \hat{P}_\phi/\partial e)-(\partial E/\partial e)(\partial \hat{P}_\phi/\partial p)},\\
\dot{p}&=&\frac{(\partial \hat{P}_\phi/\partial e)\dot{E}-(\partial E/\partial e)(\dot{P_\phi}/\mu)}{(\partial E/\partial p)(\partial \hat{P}_\phi/\partial e)-(\partial E/\partial e)(\partial \hat{P}_\phi/\partial p)}\,.
\eea
\ees
We fully expand these partial derivatives above for convenience
\begin{widetext}
\bea
\frac{\partial E}{\partial p}&=&\frac{a_1 \left(F_3-\frac{F_1 F_5}{F_2}\right)}{2 F_2 \sqrt{\frac{F_1}{F_2}}}+\frac{b_1 \left(f_3 F_1+c_1 F_3+-\frac{c_1 F_1 F_5}{F_2}\right)}{2 F_2 \sqrt{\frac{c_1 F_1}{F_2}+1}}+f_1 \sqrt{\frac{F_1}{F_2}}+f_2 \sqrt{\frac{c_1 F_1}{F_2}+1},\\
\frac{\partial E}{\partial e}&=&\frac{a_1 \left(F_4-\frac{F_1 F_6}{F_2}\right)}{2 F_2 \sqrt{\frac{F_1}{F_2}}}+\frac{b_1 \left(f_6 F_1+c_1 F_4-\frac{c_1 F_1 F_6}{F_2}\right)}{2 F_2 \sqrt{\frac{c_1 F_1}{F_2}+1}}+f_4 \sqrt{\frac{F_1}{F_2}}+f_5 \sqrt{\frac{c_1 F_1}{F_2}+1},\\
\frac{\partial \hat{P}_\phi}{\partial p}&=&\frac{F_3-\frac{F_1 F_5}{F_2}}{2 F_2 \sqrt{\frac{F_1}{F_2}}},\\
\frac{\partial \hat{P}_\phi}{\partial e}&=&\frac{F_4-\frac{F_1 F_6}{F_2}}{2 F_2 \sqrt{\frac{F_1}{F_2}}}\,.\label{partial}
\eea
\end{widetext}
The explicit forms of the functions $F_1\!-\!F_6$ and $f_1\!-\!f_{20}$ are given in the Appendix~\ref{sec:functions}. The evolution for auxiliary phase of radial motion now can be calculated by 
\be
\label{eq:xidotgen}
\dot \xi=\frac{(1+e\cos\xi)^2}{epM\sin\xi}\dot r+\frac{\cot\xi}{e}\dot e-\frac{1+e\cos\xi}{ep\sin\xi}\dot p \,.
\ee

\begin{figure}
\centering  
\subfigure{\includegraphics[width=0.45\columnwidth]{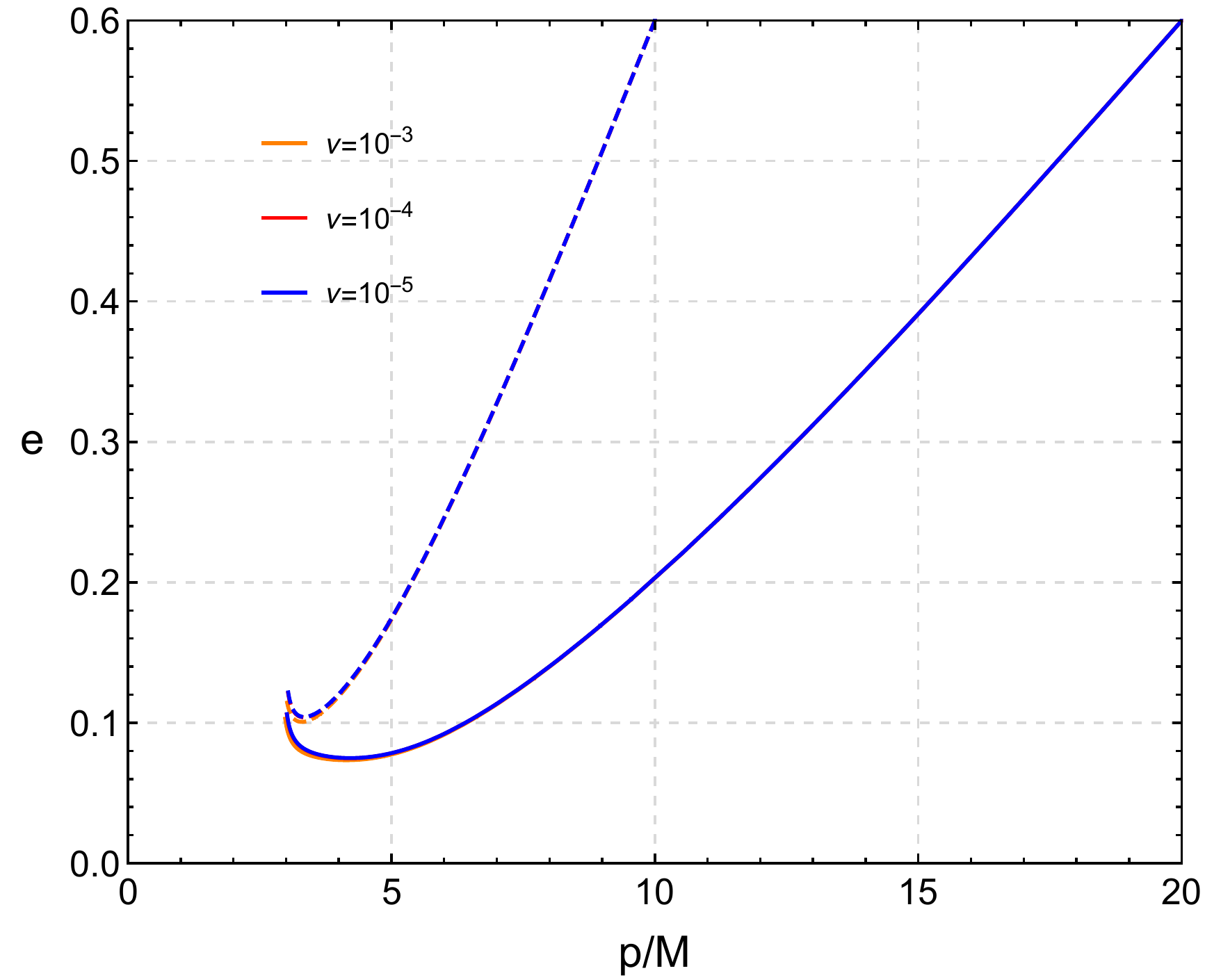}}
\subfigure{\includegraphics[width=0.45\columnwidth]{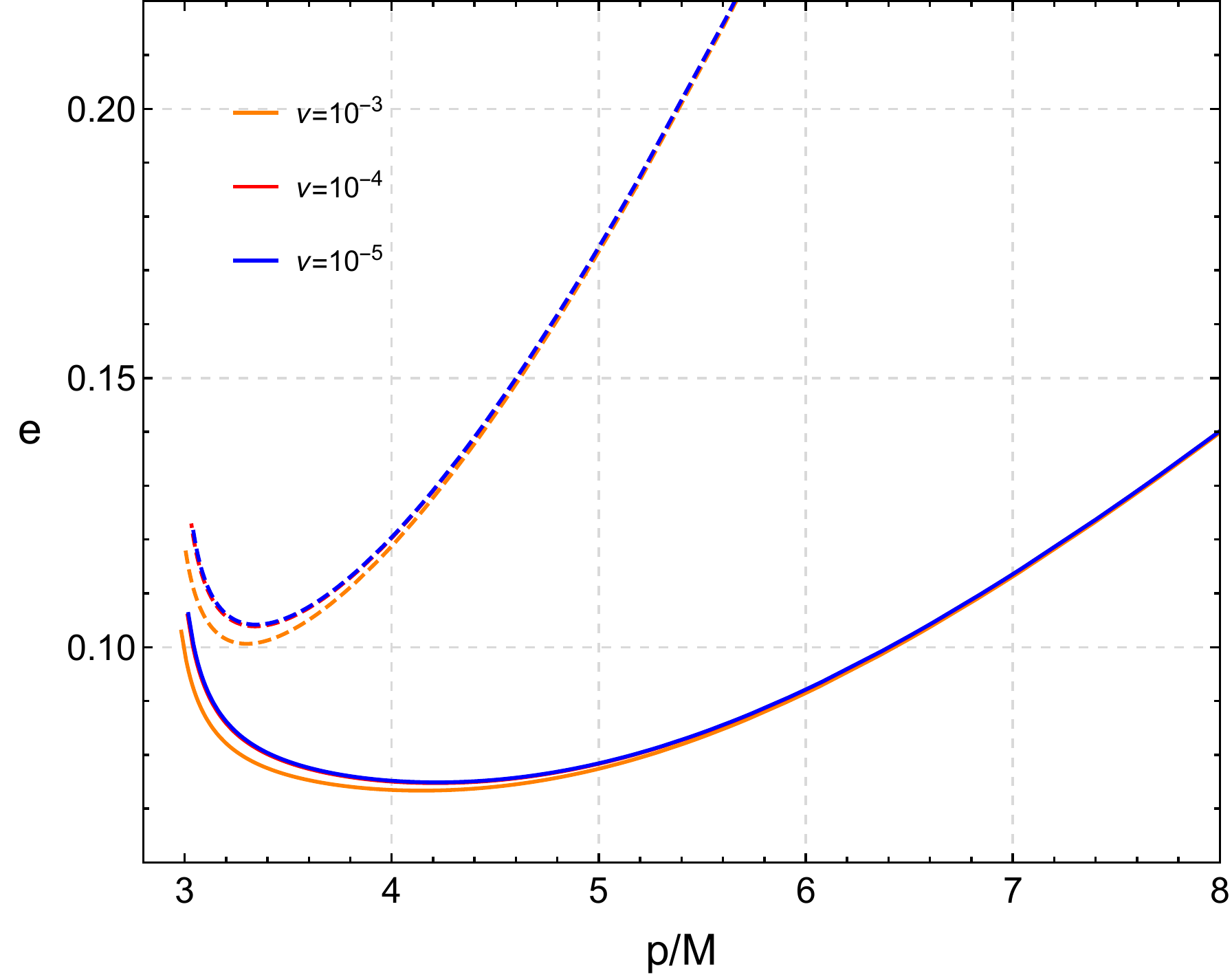}}
\caption{\emph{Illustration of orbital evolution with different mass-ratios and orbital parameters with 2PN fluxes}. The solid line represent the evolution with initial parameters $p=20M$, $e=0.6$ and the dashed one $p=10M$, $e=0.6$. The different colors mean different mass-ratios.}
\label{fig:epevolution}
\end{figure}

Figure~\ref{fig:epevolution} demonstrates the evolution of eccentricity and semilatus rectum of with different orbital parameters and mass-ratios. We can find that the evolution of $\nu = 10^{-3}$ deviate the other two evolution with smaller mass-ratio when the evolution is close to the end.

\begin{figure}
\centering  
\subfigure{\includegraphics[width=0.45\columnwidth]{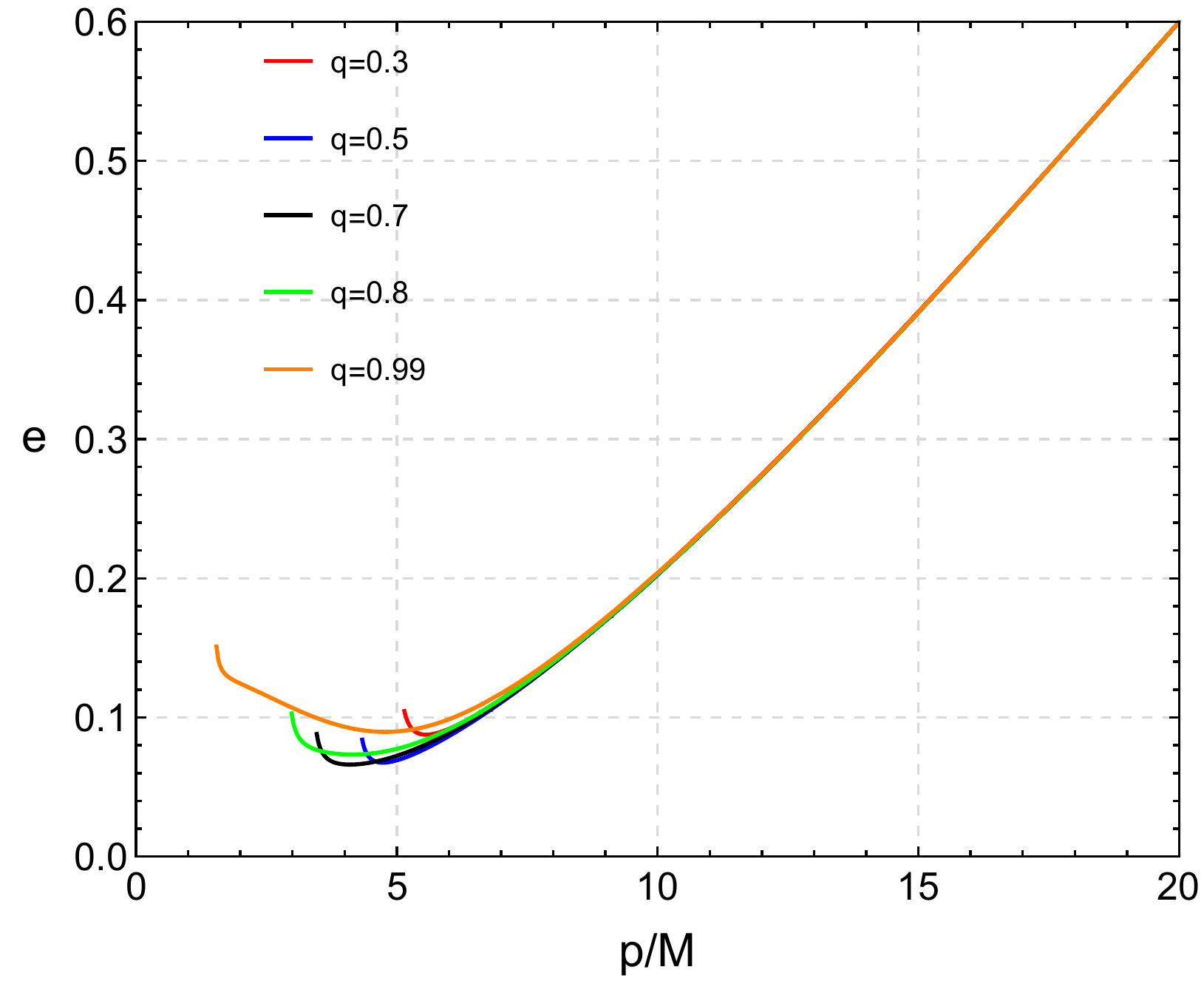}}
\subfigure{\includegraphics[width=0.45\columnwidth]{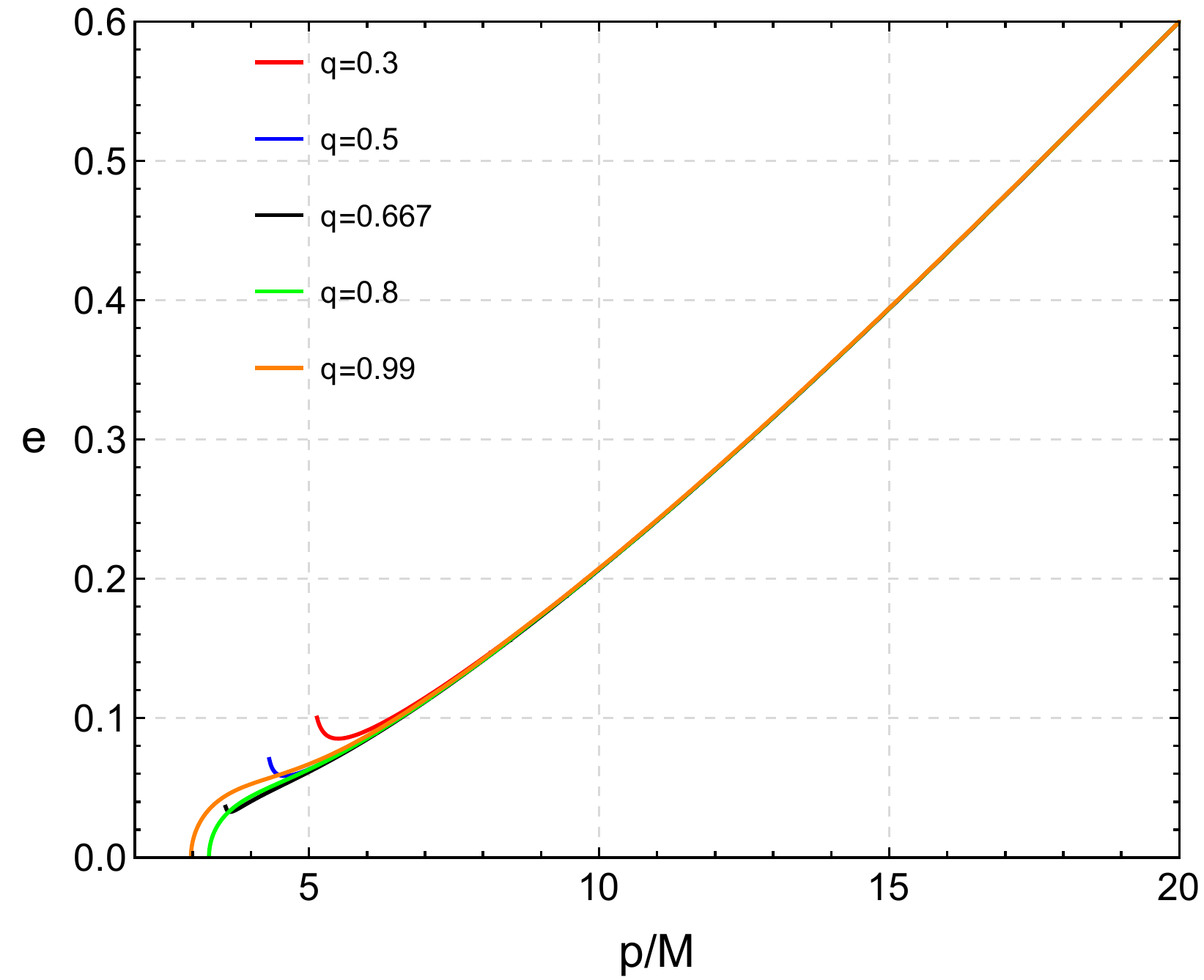}}
\caption{\emph{2PN and 3PN evolution for different spin with $\nu=0.001$}. The left and right panels show the evolution of eccentricity and semilatus rectum for five inspirals with various spins using the fluxes truncated at 2PN and 3PN respectively.}
\label{fig:2PN and 3PN envolution}
\end{figure}

Figure~\ref{fig:2PN and 3PN envolution} illustrates the evolution of orbits with 2PN (left panels) and 3PN fluxes (right panels). It looks like that the results of 3PN are not very reasonable when the orbits are extremely relativistic.

\begin{figure}
\includegraphics[width=0.9\columnwidth]{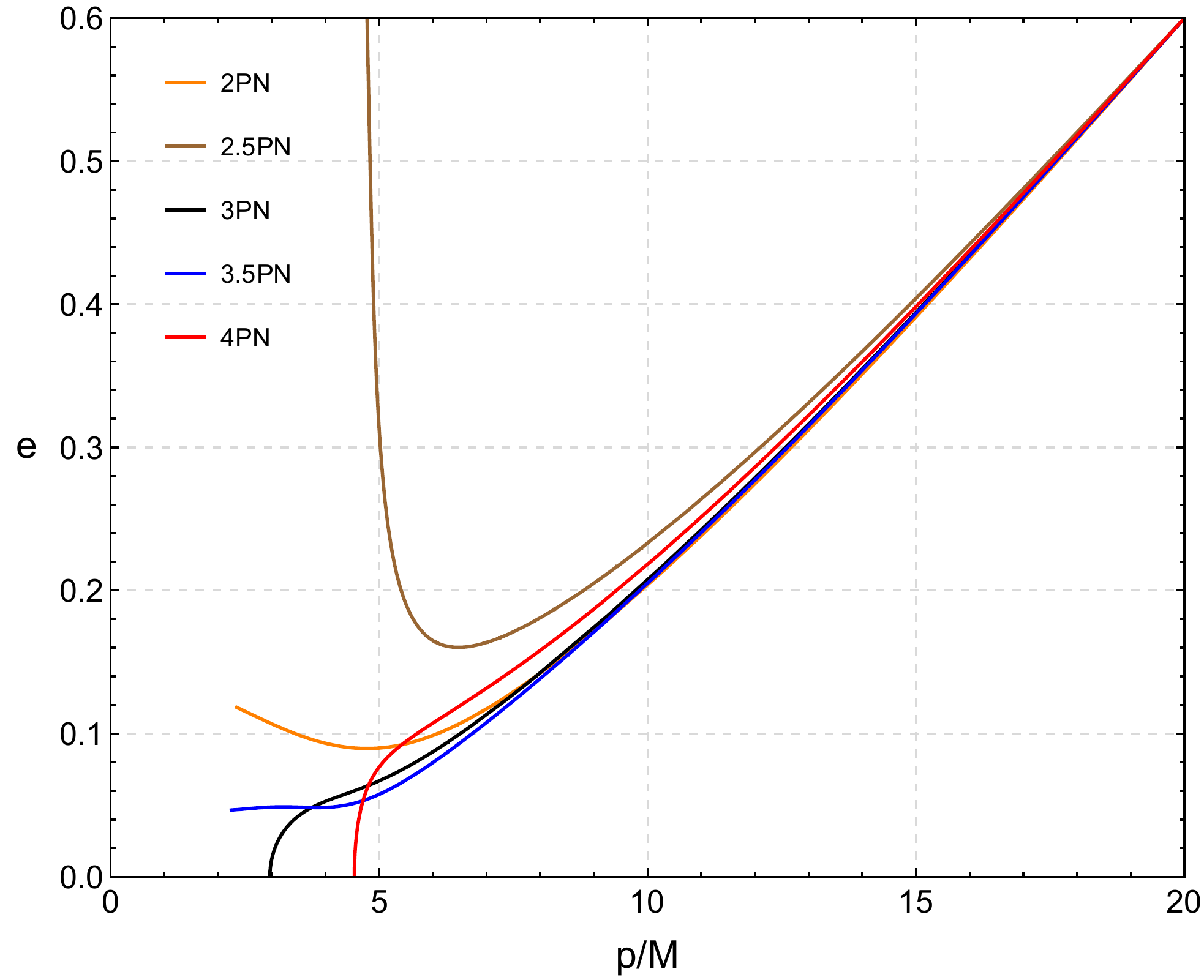}
\caption{\emph{Demonstrating the performances of different PN order  fluxes}. The figure shows inspirals with initial parameters $p=20M$, $e=0.6$, $\nu=0.001$ and spin $a=0.99M$ by using the fluxes truncated at $2$PN, $2.5$PN, $3$PN, $3.5$PN and $4$PN.}
\label{fig:different PN orders}
\end{figure}

Figure~\ref{fig:different PN orders} shows the performances of various PN fluxes on orbital evolution with very extreme spin. Similar with \cite{gair2006improved}, the 2PN formalism may be still the best choice. Therefore, we will use 2PN fluxes to calculate the orbital evolution. 

%%%%%%%%%%%%%%%%%%%%%%%%%%%%%%%%%%%%%%%%%%%%%%%%%%%%%%
\subsection{Waveform}
\label{sec:waveform}
%%%%%%%%%%%%%%%%%%%%%%%%%%%%%%%%%
In this subsection, we calculate the waveforms by solving the Teukolsky equations \cite{Teukolsky}. Our method is based on frequency-domain decomposition, and has been developed in previous works \cite{han2010gravitational,han2011constructing,han2014gravitational,han2017excitation}, in which the gravitational waveform from an eccentric EMRI with total mass $M$  at distance $R$ , latitude angle $\Theta$ and azimuthal angle $\Phi$ of could be written as 
\begin{align}
h_{+} - ih_{\times} = \frac{2}{R} \sum\limits_{lmk} \frac{Z^{\rm H}_{lmk}}{\omega^2_{mk}} {_{-2}S}^{a \omega }_{lmk}\left(\Theta\right){e}^{-i\phi_{mk}+i m \Phi},\label{eqn:GW_waveform}  \end{align}
where $l,~m,~k$ are the harmonic numbers,  $\phi_{mk} \equiv \int\omega_{mk} (t) dt$,  $_{-2}S^{a \omega}_{lmk}\left(\Theta\right)$ denotes spin-weighted spheroidal harmonics which depend on the polar angles $\Theta$ of the observer's direction of sight and the direction of orbital angular momentum of the source. $Z^{\rm H}_{lmk}$ describes the amplitude of each mode, which could be calculated by the radial component of the Teukolsky equation (see Appendix~\ref{sec:Teukolsky} for details). In this article, we set $\Theta = 0$ (``face on'') and $\Phi=0$ without losing the generality, and $\omega_{mk}$ is 
\begin{align}
\omega_{mk} = m\omega_{\phi} + k\omega_{r} ,\label{eqn:graviton_omega}      \end{align} 
where $\omega_{r}$ and $\omega_{\phi}$ denote the orbital frequencies of radial and azimuthal direction respectively which are given in Eqs. (\ref{eq:omegardef},\ref{eq:omegaphidef}). Due to our analytical solution of orbits and frequencies in the previous section, the calculation of Teukolsky-base waveform becomes very convenient and accurate. 

\begin{figure}
\centering
\subfigure[\;$p_0=10,\;e_0=0.60$]{\includegraphics[width=0.45\linewidth]{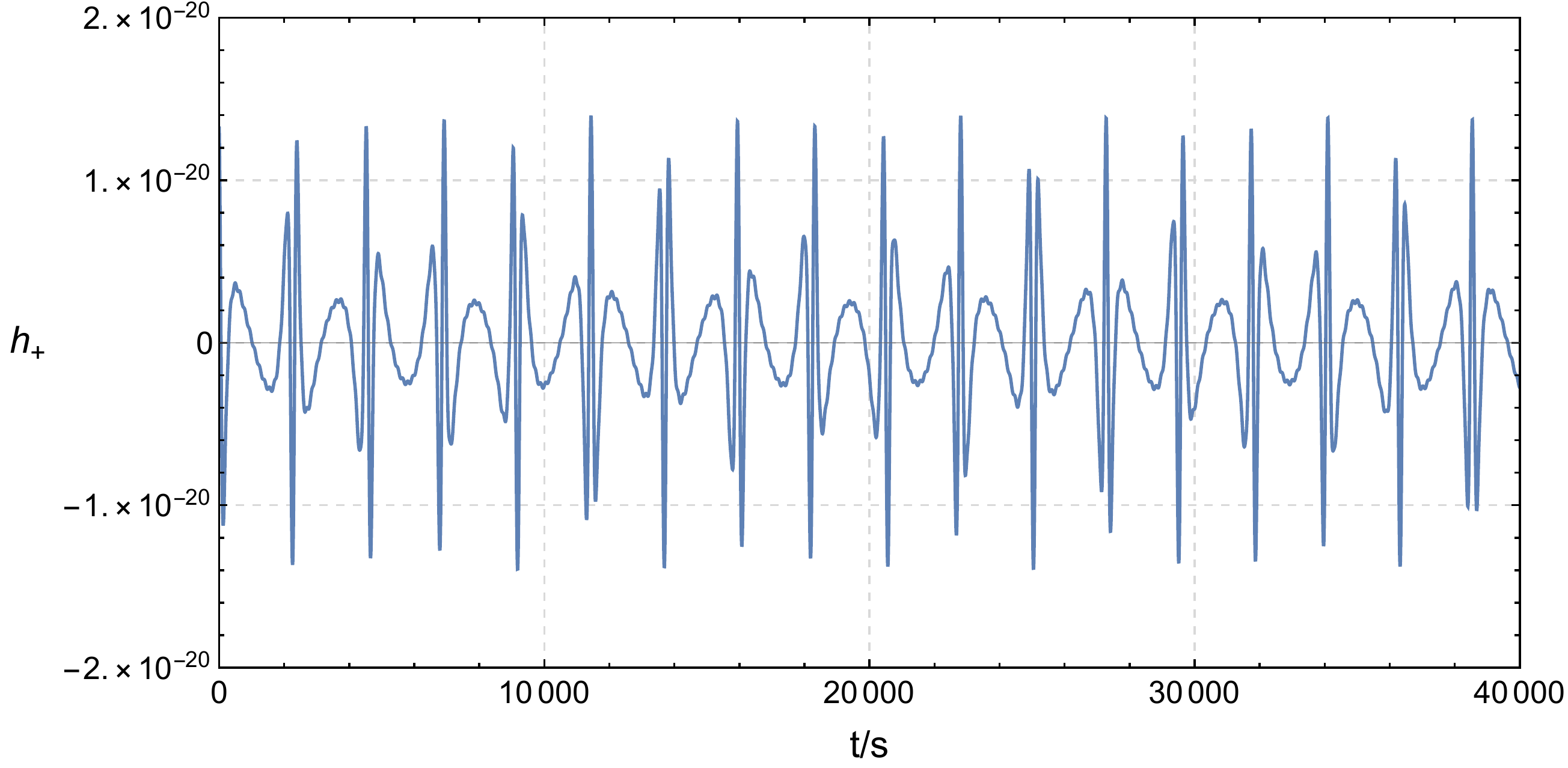}}
\quad
\subfigure[\;$p_1=7,\;e_1=0.343$]{\includegraphics[width=0.45\linewidth]{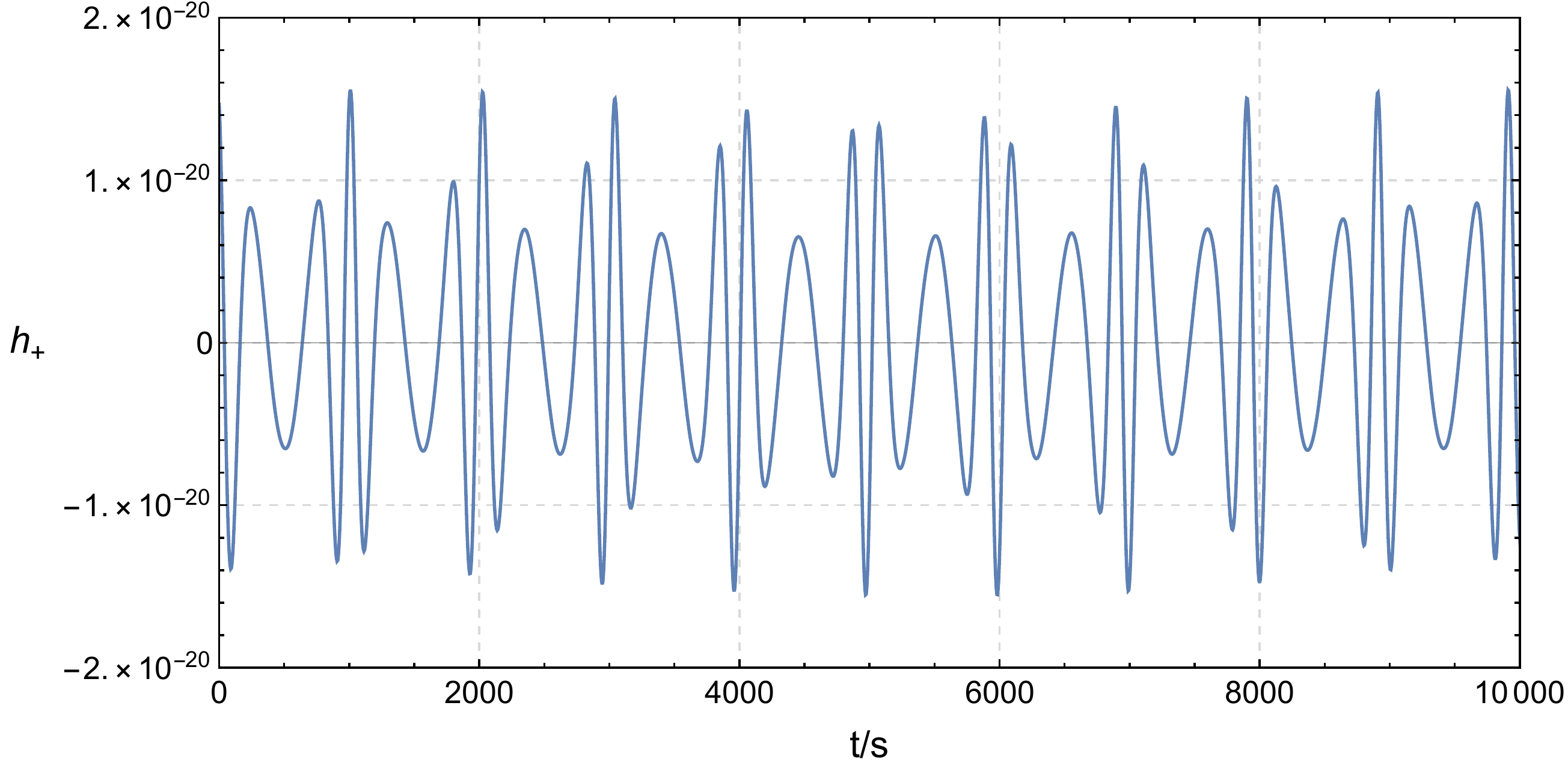}}
\quad
\subfigure[\;$ p_2=5,\;e_2=0.187$]{\includegraphics[width=0.45\linewidth]{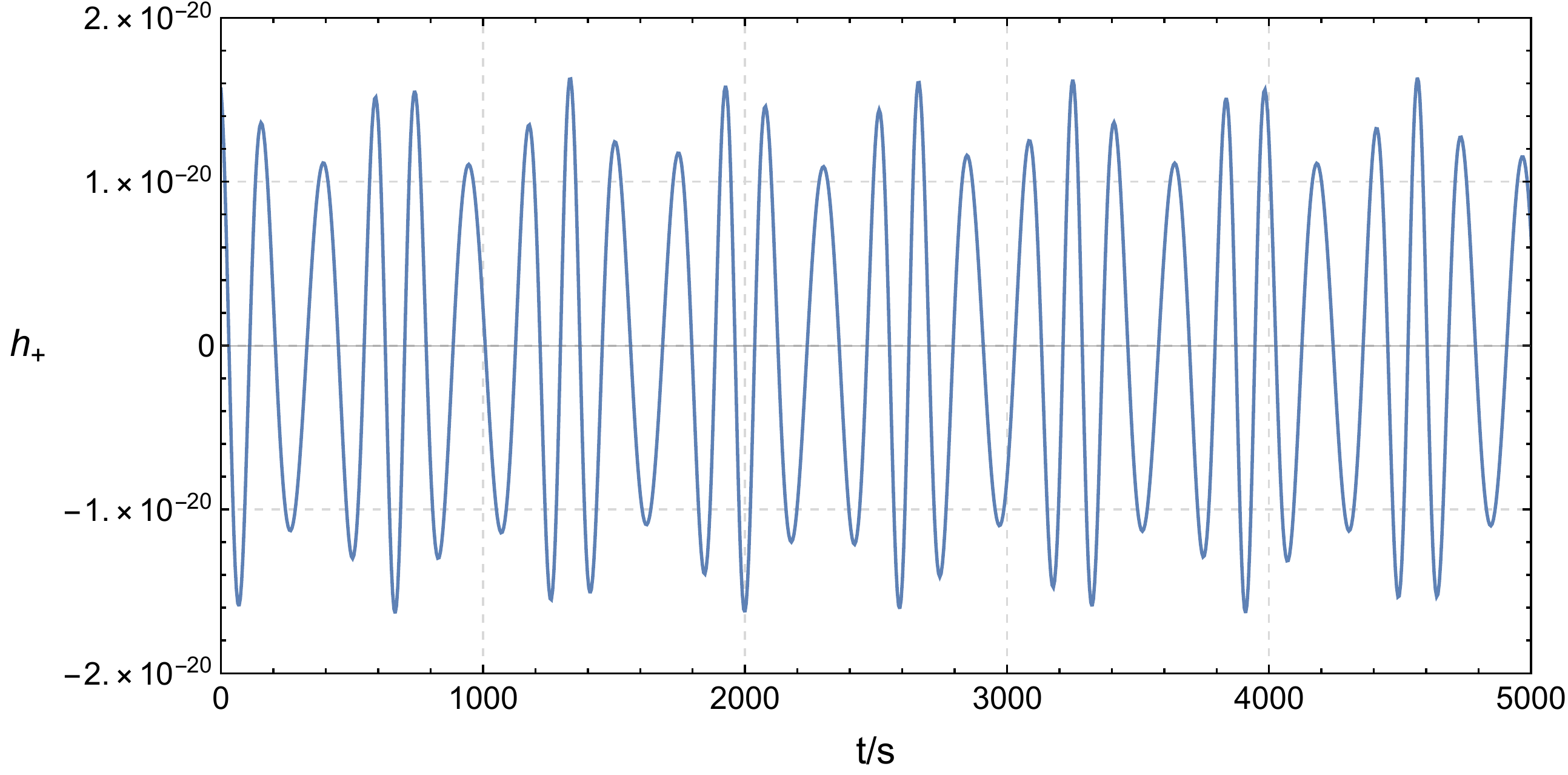}}
\quad
\subfigure[\;$ p_3=4,\;e_3=0.13$]{\includegraphics[width=0.45\linewidth]{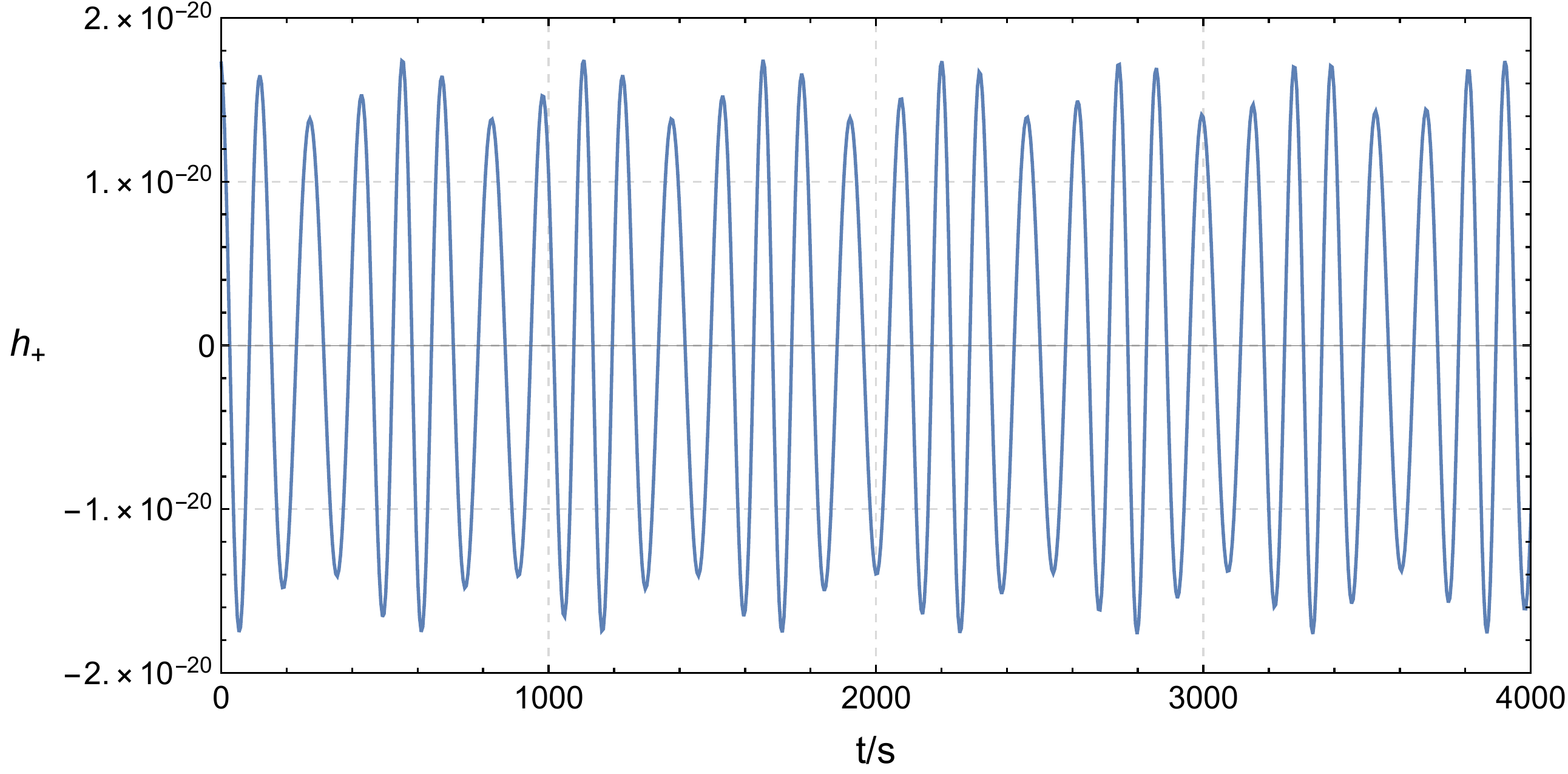}}
\caption{\emph{Four segmentations of inspiral waveforms}. These four waveforms describe the same inspiraling orbit which initial parameters are $a=0.9, p_0=10~M, e_0=0.6$ at four moments. The mass of SMBH is $10^6 m_\odot$, mass-ratio is $10^{-3}$, distance is 1 Gpc and face-on to the observer.}
\label{fig:waveform}
\end{figure}

As an example, figure~\ref{fig:waveform} illustrates the numerical waveforms of four evolution stages of an EMRI with inital parameters $p_0 = 10~M, ~e_0 = 0.6$ and mass-ratio $\nu = 10^{-3}$. We can find that as long as the time elapses, due to the semilatus rectum $p$ becomes smaller, the GW frequency becomes higher. At the same time, because of the decreasing of eccentricity $e$, the GW strains no longer vary strongly when the small object passing through the periastron and apastron. 

\begin{figure}
\includegraphics[width=0.9\columnwidth]{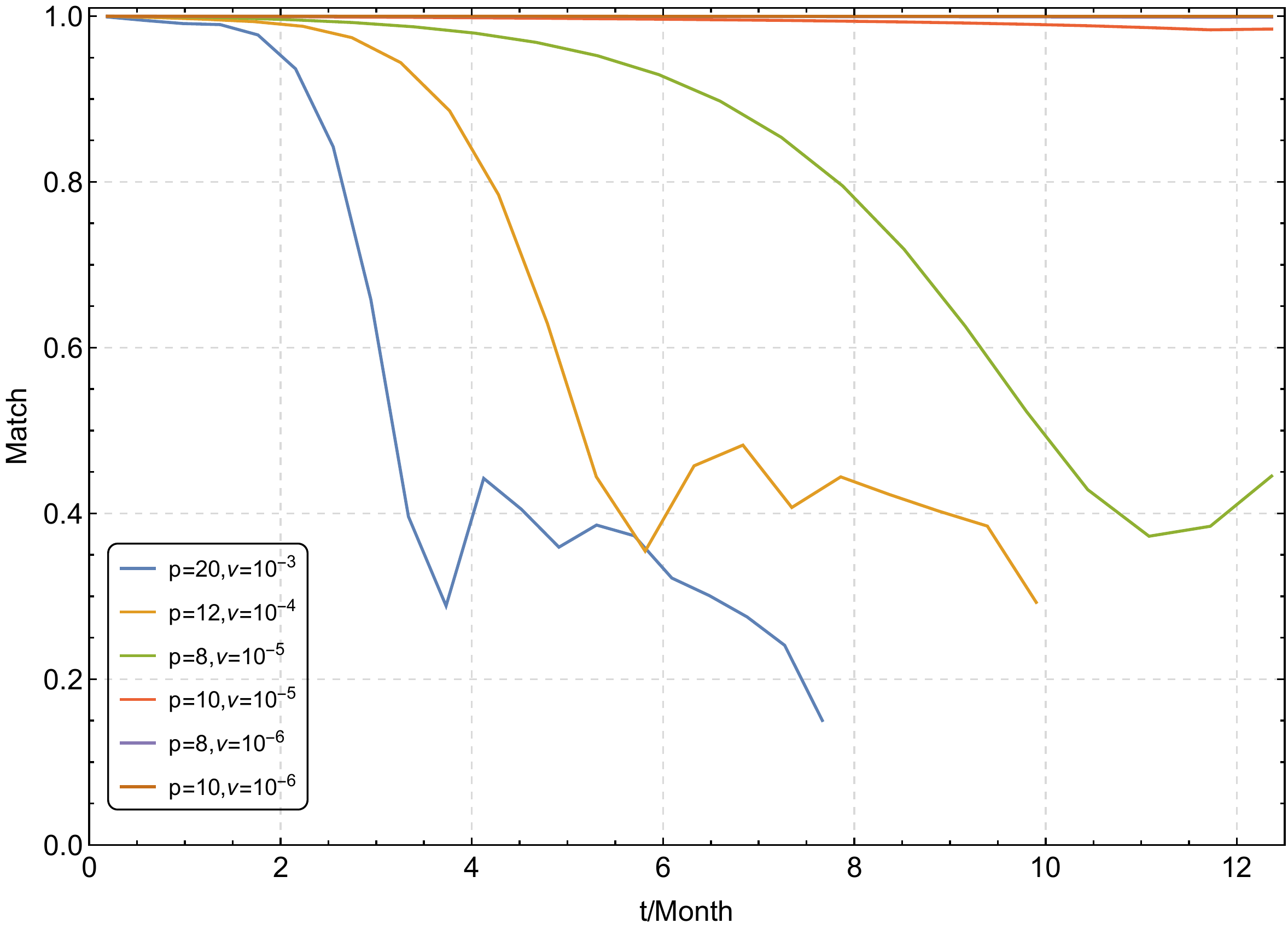}
\caption{\emph{Match of waveforms generated by test particle model and our EOB formalism.}  For all waveforms, $a=0.9M$ and $e=0.6$.}
\label{fig:match}
\end{figure}

Matched filtering \cite{finn1992detection} is widely used in GW detection in LIGO and  Virgo data analysis and also will be used in the future space-borne detectors. We employ this technology to quantitatively analyze the influence of mass-ratio on the EMRI waveforms. Figure~\ref{fig:match} illustrates the match results of the waveforms with mass-ratio correction both in conservative dynamics and fluxes and the waveforms by test particle model. We employ six EMRIs with mass-ratio from $10^{-6}$ to $10^{-3}$. The results show that for mass-ratio as low as $10^{-6}$, the test particle model may be still valid in waveform template calculations. However, for mass-ratio $\sim 10^{-5}$, if the small body at the extremely relativistic orbit around the central BH, the waveform templates of test-particle approximation will be invalid after a few of month's evolution. For mass-ratio $\sim 10^{-3}$, even for large separation of orbit, the match of two waveforms drops just after two months. We may conclude that the mass-ratio correction in EMRI waveform model should be important in the detection of such kind of systems.  

%%%%%%%%%%%%%%%%%%%%%%%%%%%%%%%%%%%%%%%
\section{Conclusions and Outlook}
\label{sec:conclusion}
%%%%%%%%%%%%%%%%%%%%%%%%%%%%%%%%%%%%%%%%

In the present paper, based on the EOB theory, we give analytical orbital solutions of elliptic EMRIs with spinning black holes. The solutions are derived with geometric parameters $p$ and $e$ instead of the EOB coordinates and momenta. The fundamental properties of the motion due to mass ratio and black hole's spin are discussed. We also give the expressions for two orbital frequencies. With these formalism at hand, it is convenient to combine with the frequency-domain Teukolsky equation, and generate accurate numerical waveforms. In addition, we express the forms of orbital evolution under gravitational radiation. We also insert the 1PN mass-ratio correction into the energy and angular momentum fluxes and the performance of fluxes with different PN orders on orbital evolution are shown.

Using matched filtering, we reveal the influence of mass ratio on the detection of EMRI GWs. We indicate that for mass-ratio $\nu \gtrsim 10^{-5}$, the conservative gravitational self-force of small objects should be considered into the construction of EMRI waveform templates. Considering the main EMRI waveform models like AK, AAK, NK and XSPEG etc. take the small objects as test particles, our model may make progress on the development of EMRI templates. 

As we mentioned before, in the present model, we temporarily omit the effective spin of the small object. In the EOB theory, this spin of effective test particle is $\sim \mu a /M$ even if the small object does not really rotate.  This is why we state that our present model only works for EMRIs and still is an improvement comparing to test particle approximation.  Furthermore, the present model can not use to inclined orbits. We will solve these two problems in the next work.    

One of the scientific targets of EMRIs is to detect the spacetime geometry of the SMBH. For this target, an accurate and efficient waveform template is needed. However, this is still a challenge now. The analytical orbital solution including mass-ratio and eccentricity given in this paper is more accurate and efficient description of the EMRI orbits. The combination of the analytical orbit and the Teukolsky equation can generate accurate waveforms. We hope our work is useful to the development of EMRI waveform template for space-borne detectors. 

%%%%%%%%%%%%%%%%%%%%%%%%%%%%%%%%%%%%%%%%
\acknowledgments
%%%%%%%%%%%%%%%%%%%%%%%%%%%%%%%%%%%%%%%%%%%%%%%%%%%%%%%%%%%%%%%%%%%%%%%%%%%%%%%%%%%%%%%%%%%%%%%%%%%%
This work is supported by NSFC No. 11773059, and we also appreciate the anonymous Referee’s suggestions about our work. This work was also supported by MEXT, JSPS Leading-edge Research Infrastructure Program, JSPS Grant-in-Aid for Specially Promoted Research 26000005, JSPS Grant-in-Aid for Scientific Research on Innovative Areas 2905: JP17H06358, JP17H06361 and JP17H06364, JSPS Core-to-Core Program A. Advanced Research Networks, JSPS Grant-in-Aid for Scientific Research (S) 17H06133, the joint research program of the Institute for Cosmic Ray Research, University of Tokyo, and by Key Research Program of Frontier Sciences, CAS, No. QYZDB-SSW-SYS016.

\appendix

\appendix

\section{log-resummed, calibrated versions of the potential}
\label{sec:potential}

The log-resummed, calibrated A-potential is given by the expression from APPENDIX A of~\cite{steinhoff2016Apotential}

\begin{widetext}

\be
A=\bar{\Delta}_u \left(\Delta _0 \nu +\nu 
   \log \left(\Delta_5 u^5+\Delta_4
   u^4+\Delta_3 u^3+\Delta_2
   u^2+\Delta_1 u+1\right)+1\right),
   \ee
with
\bes
\bea
\bar{\Delta}_u&=&\frac{1}{(K \nu -1)^2}+\frac{2u}{K \nu -1},\\
\Delta_5&=& (K \nu -1)^2 \bigg[ \frac{64}{5} \log  (u)
  + \left(-\frac{1}{3} a^2
  \left(\Delta_1^3-3\Delta_1
  \Delta_2+3 \Delta _3\right)+\frac{\Delta_1^4-4 \Delta _1^2\Delta_2+4\Delta_1\Delta_3+2\Delta_2^2-4 \Delta_4}{2 K \nu  -2}\right. \nonumber \\
  &&\left.-\frac{\Delta_1^5-5\Delta_1^3
  \Delta_2+5\Delta_1^2 \Delta_3+5\Delta_1\Delta_2^2-5
  \Delta_2 \Delta_3-5\Delta_4\Delta_1}{5 (K \nu -1)^2}+\frac{2275 \pi^2}{512}+\frac{128 \gamma }{5}-\frac{4237}{60}+\frac{256
  \log (2)}{5}\right) \bigg] , \\
\Delta_4&=&\frac{1}{96} \bigg[8 \left(6 a^2 \left(\Delta_1^2-2 \Delta_2\right) (K \nu -1)^2+3 \Delta_1^4+\Delta_1^3 (8-8 K \nu )-12 \Delta_1^2\Delta_2+12\Delta_1 (2
  \Delta_2 K \nu -2 \Delta_2+\Delta_3)\right)\nonumber\\
  &&+48\Delta_2^2-64 (K \nu -1) (3 \Delta_3-47 K \nu +47)-123 \pi ^2 (K\nu -1)^2\bigg],\\
\Delta_3&=&-a^2\Delta_1 (K \nu -1)^2-\frac{\Delta_1^3}{3}+\Delta_1^2 (K \nu -1)+\Delta_1\Delta_2-2 (K \nu -1) (\Delta_2-K
   \nu +1),\\
\Delta_2 &=& \frac{1}{2} \left(\Delta_1 (\Delta_1-4 K
   \nu +4)-2 a^2\Delta_0 (K \nu -1)^2\right),\\
\Delta_1 &=&-2 (\Delta_0+K) (K \nu -1),\\
\Delta_0&=&K (K \nu -2),
\eea
where $K$ is a calibration parameter tuned to numerical-relativity simulations whose most recently updated value was determined in Eq.~(4.8) of Ref.~\cite{bohe2017improved}
\bea
K=267.788 \nu ^3-126.687 \nu ^2+10.2573 \nu +1.7336
\eea
\ees
\end{widetext}

The $D$-potential is 
\bes
\bea
D_{\rm Taylor}&=&1+6\nu u^2+2 \nu u^3(26-3\nu)\nonumber\\
\frac{-1}{D}(u)&=&1+\log\left[D_{\rm Taylor}\right].
\eea
\ees

\section{functions that appear in the solution for $\dot{e}$ and $\dot{p}$}
\label{sec:functions}

The quantities repeated 
\begin{widetext}
\bes
\bea
F_1&=&2 X Y_4 \sqrt{X^2-Y_2 Z}+X^2 Y_3-Y Y_2, \\
F_2&=&X^4-2 X^2 Y_1+Y^2, \\
F_3&=&\frac{X Y_4 \left(2 f_7 X-f_{13} Y_2-f_{10} Z\right)}{\sqrt{X^2-Y_2 Z}}+2 \left(f_{12} X+f_7 Y_4\right) \sqrt{X^2-Y_2 Z}+f_{11} X^2+2 f_7 X Y_3-f_8 Y_2-f_{10} Y, \\
F_4&=&\frac{X Y_4 \left(2 f_{14} X-f_{20} Y_2-f_{17} Z\right)}{\sqrt{X^2-Y_2 Z}}+2 \left(f_{19} X+f_{14} Y_4\right) \sqrt{X^2-Y_2 Z}+f_{18} X^2+2 f_{14} X Y_3-f_{15} Y_2-f_{17} Y, \\
F_5&=&4 f_7 \left(X^3-X Y_1\right)-2 f_9 X^2+2 f_8 Y, \\
F_6&=&4 f_{14} \left(X^3-X Y_1\right)-2 f_{16} X^2+2 f_{15} Y,
\eea
\ees
\bes
\bea
f_1&=&\frac{a_1 \left(3 a^2 (A-2) (1-e)^2 p^2+(1-e)^2 j_5 p^3-5 p^4\right)+4 a (1-e)^3 p}{p^5-a^2 (A-2) (1-e)^2 p^3}, \\
f_2&=&\frac{1}{2 b_1}\bigg[j_1-\frac{c_1^2 \left(2 p-(1-e)^2 j_5\right) \left(a^2 (1-e)^2+A p^2\right)}{(1-e)^4}\bigg], \\
f_3&=&-\frac{c_1^2 \left(2 p-(1-e)^2 j_5\right)}{(1-e)^2}, \\
f_4&=&\frac{3 a_1}{1-e}-\frac{a_1 (1-e) \left(2 a^2 (A-2) p^3-j_5 p^4\right)+2 a \left(20 a^2-10\right) (1-e)^4 \nu }{p^5-a^2 (A-2) (1-e)^2 p^3}, \\
f_5&=&\frac{c_1}{2 b_1 (1-e)^3}\bigg[j_3+c_1 \left(a^2 (1-e)^2+A p^2\right) \left(j_5 p-2 a^2 (A-2)\right)\bigg], \\
f_6&=&\frac{c_1^2 \left(j_5 p-2 a^2 (A-2)\right)-2 c_1}{1-e}, \\
f_7&=&f_1-\frac{4 a (1+e)^3 p-a_2 \left(-3 a^2 (B-2) (1+e)^2 p^2+(1+e)^2 j_6 p^3+5 p^4\right)}{p^5-a^2 (B-2) (1+e)^2 p^3}, \\
f_8&=&c_1 j_1-c_2 j_2-\frac{2 c_1^3 \left(2 p-(1-e)^2 j_5\right) \left(a^2 (1-e)^2+A p^2\right)}{(1-e)^4}+\frac{2 c_2^3 \left(2 p+(1+e)^2 j_6\right) \left(a^2 (1+e)^2+B p^2\right)}{(1+e)^4}, \\
f_9&=&c_1 j_1+c_2 j_2-\frac{2 c_1^3 \left(2 p-(1-e)^2 j_5\right) \left(a^2 (1-e)^2+A p^2\right)}{(1-e)^4}-\frac{2 c_2^3 \left(2 p+(1+e)^2 j_6\right) \left(a^2 (1+e)^2+B p^2\right)}{(1+e)^4}, \\
f_{10}&=&j_1-j_2-\frac{c_1^2 \left(2 p-(1-e)^2 j_5\right) \left(a^2 (1-e)^2+A p^2\right)}{(1-e)^4}+\frac{c_2^2 \left(2 p+(1+e)^2 j_6\right) \left(a^2 (1+e)^2+B p^2\right)}{(1+e)^4}, \\
f_{11}&=&j_1+j_2-\frac{c_1^2 \left(2 p-(1-e)^2 j_5\right) \left(a^2 (1-e)^2+A p^2\right)}{(1-e)^4}-\frac{c_2^2 \left(2 p+(1+e)^2 j_6\right) \left(a^2 (1+e)^2+B p^2\right)}{(1+e)^4}, \\
f_{12}&=&b_2 f_2+\frac{b_1}{2 b_2}\bigg[j_2-\frac{c_2^2 \left(2 p+(1+e)^2 j_6\right) \left(a^2 (1+e)^2+B p^2\right)}{(1+e)^4}\bigg], \\
f_{13}&=&f_3+\frac{c_2^2 \left(2 p+(1+e)^2 j_6\right)}{(1+e)^2}, \\
f_{14}&=&f_4-\frac{3 a_2}{1+e}-\frac{a_2 (1+e) \left(2 a^2 (B-2) p^3+j_6 p^4\right)+2 a \left(20 a^2-10\right) (1+e)^4 \nu }{p^5-a^2 (B-2) (1+e)^2 p^3}, \\
f_{15}&=&\frac{c_1^2 \bigg[2 b_1^2 \left(j_5 p-2 a^2 (A-2)\right)+\frac{j_3-2 \left(a^2 (1-e)^2+A p^2\right)}{(1-e)^2}\bigg]}{1-e}-\frac{c_2^2 \bigg[2 b_2^2 \left(2 a^2 (B-2)+j_6 p\right)+\frac{j_4+2 \left(a^2 (1+e)^2+B p^2\right)}{(1+e)^2}\bigg]}{1+e}, \\
f_{16}&=&\frac{c_1^2 \bigg[2 b_1^2 \left(j_5 p-2 a^2 (A-2)\right)+\frac{j_3-2 \left(a^2 (1-e)^2+A p^2\right)}{(1-e)^2}\bigg]}{1-e}+\frac{c_2^2 \bigg[2 b_2^2 \left(2 a^2 (B-2)+j_6 p\right)+\frac{j_4+2 \left(a^2 (1+e)^2+B p^2\right)}{(1+e)^2}\bigg]}{1+e}, \\
f_{17}&=&\frac{c_1^2 \left(a^2 (1-e)^2+A p^2\right) \left(j_5 p-2 a^2 (A-2)\right)+c_1 j_3}{(1-e)^3}-\frac{c_2^2 \left(a^2 (1+e)^2+B p^2\right) \left(2 a^2 (B-2)+j_6 p\right)+c_2 j_4}{(1+e)^3}, \\
f_{18}&=&\frac{c_1^2 \left(a^2 (1-e)^2+A p^2\right) \left(j_5 p-2 a^2 (A-2)\right)+c_1 j_3}{(1-e)^3}+\frac{c_2^2 \left(a^2 (1+e)^2+B p^2\right) \left(2 a^2 (B-2)+j_6 p\right)+c_2 j_4}{(1+e)^3}, \\
f_{19}&=&b_2 f_5+\frac{\left(b_1 c_2\right) \left(c_2 \left(a^2 (1+e)^2+B p^2\right) \left(2 a^2 (B-2)+j_6 p\right)+j_4\right)}{(1+e) \left(2 b_2 (1+e)^2\right)}, \\
f_{20}&=&f_6+\frac{c_2^2 \left(-2 a^2 (B-2)-j_6 p\right)-2 c_2}{1+e}, 
\eea
\ees
where
\bes
\bea
X&=&a_1-a_2, Y=b_1^2 c_1-b_2^2 c_2, Y_1=b_1^2 c_1+b_2^2 c_2, Y_2=b_1^2-b_2^2, Y_3=b_1^2+b_2^2, Y_4=b_1 b_2, Z=c_1-c_2, \\
d&=&\frac{(1-e)^5\Delta _{5}(r_1)}{p^5}+\frac{(1-e)^4\Delta _4}{p^4}+\frac{(1-e)^3\Delta _3}{p^3}+\frac{(1-e)^2\Delta _2}{p^2}+\frac{(1-e)\Delta _1}{p}+1, \\
o&=&\frac{(1+e)^5\Delta _{5}(r_2)}{p^5}+\frac{(1+e)^4\Delta _4 }{p^4}+\frac{(1+e)^3 \Delta _3}{p^3}+\frac{(1+e)^2\Delta _2}{p^2}+\frac{(1+e)\Delta _1}{p}+1, \\
A&=&A\left(r_1\right)=\Delta _u(r_1) \left(\Delta _0 \nu +\nu  \log (d)+1\right), \\
B&=&A\left(r_2\right)=\Delta _u(r_2) \left(\Delta _0 \nu +\nu  \log (o)+1\right), \\
v_1&=&\frac{2 \left(\Delta _0 \nu +\nu  \log (d)+1\right)}{K \nu -1}, \\
v_2&=&\frac{2 \left(\Delta _0 \nu +\nu  \log (o)+1\right)}{K \nu -1}, \\
j_1&=&\frac{c_1}{1-e}\bigg[\frac{2 A p}{1-e}+\frac{\frac {\partial d} {\partial e} \nu  p \Delta _{u}(r_1)}{d}-v_1\bigg], \\
j_2&=&\frac{c_2}{1+e}\bigg[\frac{2 B p}{1+e}-\frac{\nu  \frac {\partial o} {\partial e} p \Delta _{u}(r_2)}{o}-v_2\bigg], \\
j_3&=&(1-e)\bigg[\frac{\frac {\partial d} {\partial e} \nu  p^2 \Delta _{u}(r_1)}{d}-p v_1-2 a^2 (1-e)\bigg], \\
j_4&=&(1+e)\bigg[\frac{\frac {\partial o} {\partial e} \nu  p^2 \Delta _{u}(r_2)}{o}+p v_2+2 a^2 (1+e)\bigg], \\
j_5&=&\frac{a^2 (1-e)}{p}\bigg[\frac{\frac {\partial d} {\partial e} \nu \Delta _{u}(r_1)}{d}-\frac{v_1}{p}\bigg], \\
j_6&=&\frac{a^2 (1+e)}{p}\bigg[\frac{\frac {\partial o} {\partial e} \nu \Delta _{u}(r_2)}{o}+\frac{v_2}{p}\bigg],
\eea
\ees
\end{widetext}

\section{validating our results with test-particle limit and nonspinning case}
\begin{figure}
\centering  
\subfigure{\includegraphics[width=0.45\columnwidth]{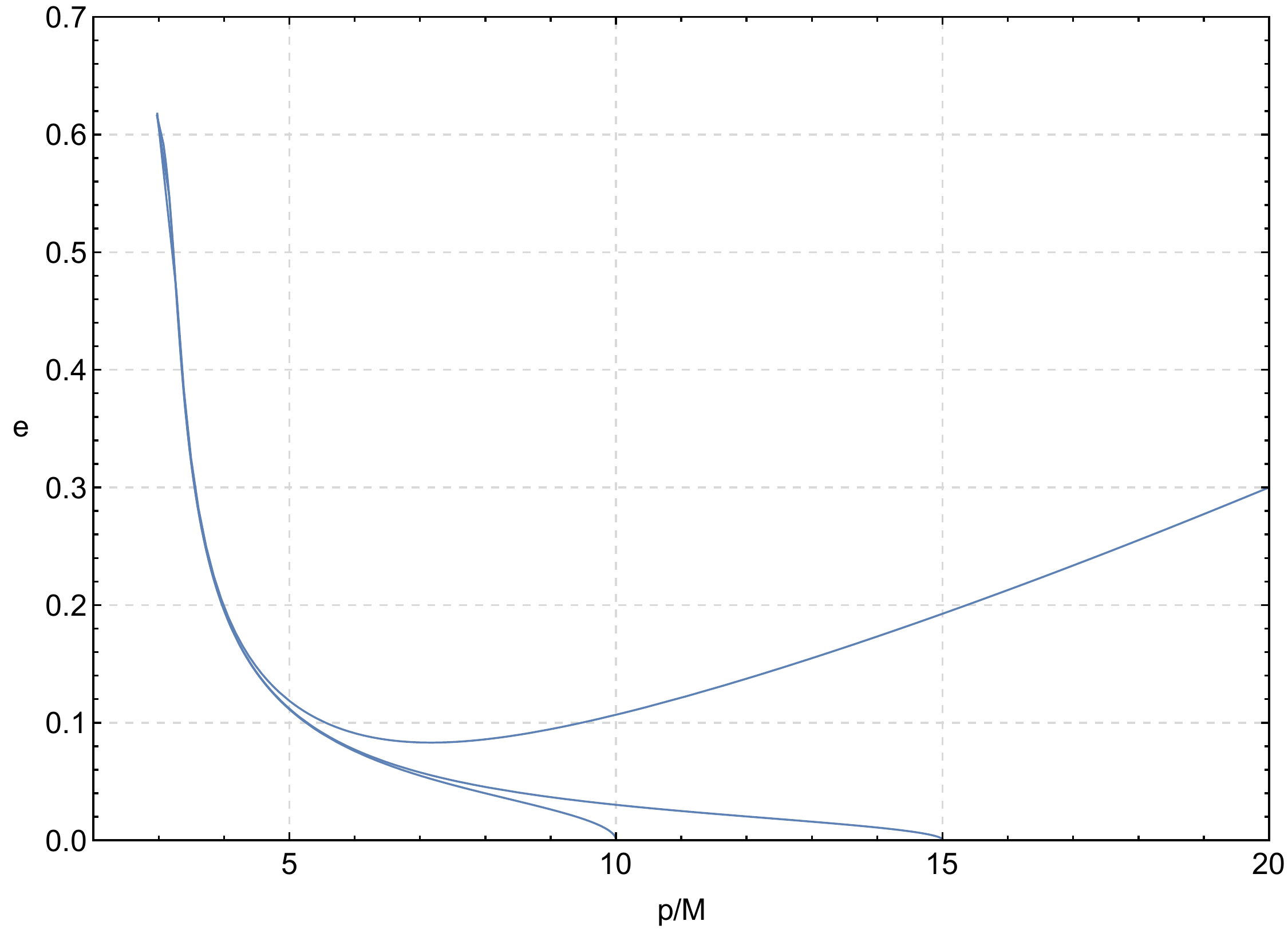}}
\subfigure{\includegraphics[width=0.45\columnwidth]{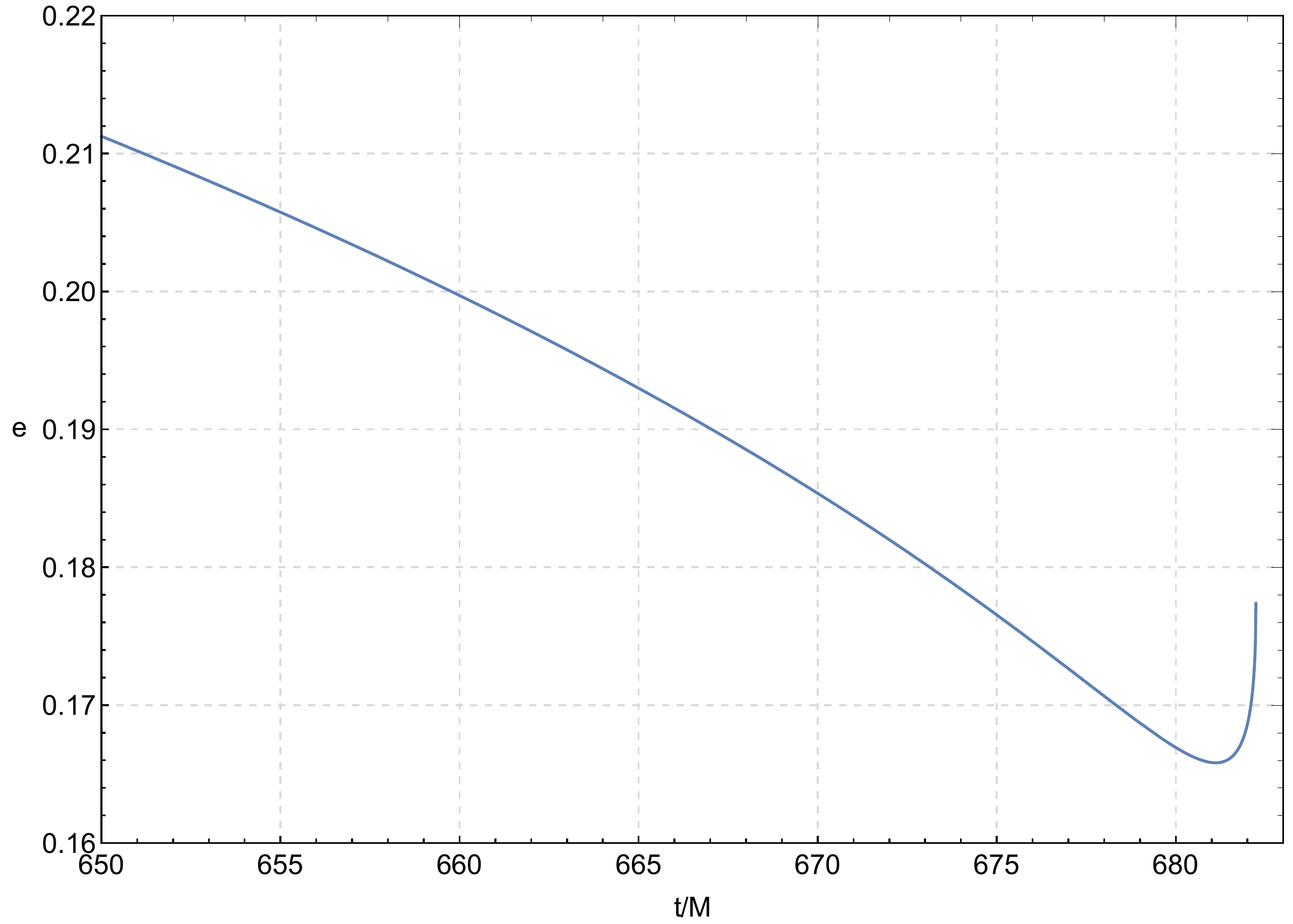}}
\caption{\emph{Validating our results with test-particle limit and nonspinning case}. The left panel is orbital evolution by taking $\nu \rightarrow 0$ in our formalism. The right panel repeats the evolution of nonspinning binaries in \cite{hinderer2017foundations}.}
\label{fig:compare}
\end{figure}

In the left panel of Figure~\ref{fig:compare}, we repeat the result in FIG.1 of~\cite{gair2006improved} by setting the mass-ratio to zero in the conservative dynamics and 1PN terms in fluxes. This proves that our Eqs. (\ref{partial}) are correct when goes back the test particle limit in Kerr spacetime. By setting the Kerr parameter $a = 0$ in our formalism, the right panel repeats the orbital evolution in FIG.6 of~\cite{hinderer2017foundations} for an equal-mass binary without spin. It shows that our formalism coincide with the Schwarzschild case when $a = 0$ but with mass-ratio $\nu = 0.25$. Combining the both results, our formalism and codes for orbital evolution are validated.

\section{The Teukolsky equation}
\label{sec:Teukolsky}
The gravitational perturbation of Kerr space-time is described by the Teukolsky equation by the Weyl curvature (complex) scalar $\psi_4$, decomposed in frequency domain $\psi_4=\rho^4\int^{+\infty}_{-\infty}{d\omega\sum_{lm}{R_{lm\omega}(r)_{~-2}S^{a\omega}_{lm}(\theta)e^{im\phi}e^{-i\omega
			t}}}$ with spin-weighted spheroidal harmonics $_{~-2}S^{a\omega}_{lm}$, obeys \cite{Teukolsky}:

\begin{equation}
\Delta^2\frac{d}{dr}\left(\frac{1}{\Delta}\frac{d
	R_{lm\omega}}{dr}\right)-V(r)R_{lm\omega}=-\mathcal{T}_{lm\omega}(r),
\label{Teukolsky}
\end{equation}
where $\mathcal{T}_{lm\omega}(r)$ is the source term, which is connected by the stress-energy tensor of the perturbation source, and the potential is
\begin{equation}
V(r)=-\frac{K^2+4i(r-M)K}{\Delta}+8i\omega r+\lambda,
\end{equation}
where $K=(r^2+a^2)\omega-ma, ~\lambda=E_{lm}+a^2\omega^2-2a m w-2$ and $\Delta = r^2-2Mr+a^2$. 

First, we consider the homogeneous Teukolsky equation where the source term is zero. We can solve it by analytical expansion, as discussed in \cite{sasaki2003analytic,fujita2004new} and here we don't go into technical details of it. The homogeneous Teukolsky equation allows two independent solutions $R^{\rm H}_{lm\omega}$, which is purely ingoing at the horizon, and $R^\infty_{lm\omega}$, which is purely outgoing at infinity:
\begin{align}\nonumber
\label{RH}
R^{\rm H}_{lm\omega}&=B^{\rm hole}_{lm\omega}\Delta^2 e^{-ipr*},\quad
r\rightarrow r_+\\
R^{\rm H}_{lm\omega}&=B^{out}_{lm\omega}r^3 e^{i\omega
	r*}+r^{-1}B^{\rm in}_{lm\omega} e^{-i\omega r*},\quad r\rightarrow
\infty;
\end{align}
\begin{align}\nonumber
\label{Rinf}
R^{\infty}_{lm\omega}&=D^{\rm{out}}_{lm\omega} e^{ip r*}+\Delta^2
D^{\rm in}_{lm\omega} e^{-ip r*},\quad
r\rightarrow r_+\\
R^{\infty}_{lm\omega}&=r^3 D^{\infty}_{lm\omega} e^{i\omega
	r*},\quad r\rightarrow \infty,
\end{align}
where $p= \omega-\frac{ma}{2Mr_+}$, $r_+=M+\sqrt{M^2-a^2}$ and $r*$ is the tortoise coordinate related to $r$ by $dr*/dr = (r^2+a^2)/\Delta$

Then, using the homogeneous solutions and proper boundary conditions, we can construct the solution to radial Teukolsky equation with source term. By imposing BH boundary condition, i.e. wave being purely outgoing at infinity and purely ingoing at horizon, the radial function is:
\begin{align}
\label{BHsolution}
	R^{\rm BH}_{lm\omega}(r)=\frac{R^{\infty}_{lm\omega}(r)}{2i\omega
		B^{\rm in}_{lm\omega}D^{\infty}_{lm\omega}}\int^{r}_{r_+}{dr'\frac{R^{\rm H}_{lm\omega}(r')\mathcal{T}_{lm\omega}(r')}{\Delta(r')^2}}+ \\ \frac{R^{\rm H}_{lm\omega}(r)}{2i\omega
		B^{\rm in}_{lm\omega}D^{\infty}_{lm\omega}}\int^{\infty}_{r}{dr'\frac{R^\infty_{lm\omega}(r')\mathcal{T}_{lm\omega}(r')}{\Delta(r')^2}}
\end{align}

The asymptotic behavior of this solution near horizon and infinity is:
\begin{align}
\label{eq_RBH_asym}
R^{\rm BH}_{lm\omega}(r\rightarrow \infty)&=Z^{\rm H}_{lm\omega}r^3 e^{i\omega
	r*},\\
R^{\rm BH}_{lm\omega}(r\rightarrow r_+)&=Z^{\infty}_{lm\omega}\Delta^2 e^{-ip
	r*}.
\end{align}

By taking the limit at $r\rightarrow \infty$ and $r\rightarrow r_+$ of the solution (Eq. \ref{BHsolution}), with the asymptotic behavior of homogeneous solutions (Eq. \ref{RH}, \ref{Rinf}), one can find the amplitudes $Z^{H,\infty}_{lm\omega}$:
\begin{equation}
	Z^{\rm H}_{lm\omega} = \frac{1}{2i\omega B^{\rm in}_{lm\omega}} \int_{r_+}^{r} dr' \frac{R^{\rm H}_{lm\omega}(r') \mathcal T _{lm\omega}(r') }{\Delta(r')^2 }
\end{equation}
\begin{equation}
Z^\infty_{lm\omega} = \frac{B^{\rm H}_{lm\omega}}{2i\omega B^{\rm in}_{lm\omega} D^{\infty }_{lm\omega}} \int_{r}^{\infty} dr' \frac{R^\infty_{lm\omega}(r') \mathcal T _{lm\omega}(r') }{\Delta(r')^2 }
\end{equation}

%\bibliographystyle{unsrt}
%\bibliography{references}

\begin{thebibliography}{10}

\bibitem{abbott2016observation}
Benjamin~P Abbott, Richard Abbott, TD~Abbott, MR~Abernathy, Fausto Acernese,
  Kendall Ackley, Carl Adams, Thomas Adams, Paolo Addesso, RX~Adhikari, et~al.
\newblock Observation of gravitational waves from a binary black hole merger.
\newblock {\em Physical review letters}, 116(6):061102, 2016.

\bibitem{abbott2016gw151226}
Benjamin~P Abbott, R~Abbott, TD~Abbott, MR~Abernathy, F~Acernese, K~Ackley,
  C~Adams, T~Adams, P~Addesso, RX~Adhikari, et~al.
\newblock Gw151226: observation of gravitational waves from a 22-solar-mass
  binary black hole coalescence.
\newblock {\em Physical review letters}, 116(24):241103, 2016.

\bibitem{scientific2017gw170104}
LIGO Scientific, BP~Abbott, R~Abbott, TD~Abbott, F~Acernese, K~Ackley, C~Adams,
  T~Adams, P~Addesso, RX~Adhikari, et~al.
\newblock Gw170104: observation of a 50-solar-mass binary black hole
  coalescence at redshift 0.2.
\newblock {\em Physical Review Letters}, 118(22):221101, 2017.

\bibitem{abbott2017gw170608}
Benjamin~P Abbott, R~Abbott, TD~Abbott, F~Acernese, K~Ackley, C~Adams, T~Adams,
  P~Addesso, RX~Adhikari, VB~Adya, et~al.
\newblock Gw170608: Observation of a 19 solar-mass binary black hole
  coalescence.
\newblock {\em The Astrophysical Journal Letters}, 851(2):L35, 2017.

\bibitem{abbott2017gw170814}
Benjamin~P Abbott, R~Abbott, TD~Abbott, F~Acernese, K~Ackley, C~Adams, T~Adams,
  P~Addesso, RX~Adhikari, VB~Adya, et~al.
\newblock Gw170814: a three-detector observation of gravitational waves from a
  binary black hole coalescence.
\newblock {\em Physical review letters}, 119(14):141101, 2017.

\bibitem{abbott2017gw170817}
Benjamin~P Abbott, Rich Abbott, TD~Abbott, Fausto Acernese, Kendall Ackley,
  Carl Adams, Thomas Adams, Paolo Addesso, RX~Adhikari, VB~Adya, et~al.
\newblock Gw170817: observation of gravitational waves from a binary neutron
  star inspiral.
\newblock {\em Physical Review Letters}, 119(16):161101, 2017.

\bibitem{danzmann1996lisa}
Karsten Danzmann, LISA~Study Team, et~al.
\newblock Lisa: Laser interferometer space antenna for gravitational wave
  measurements.
\newblock {\em Classical and Quantum Gravity}, 13(11A):A247, 1996.

\bibitem{hu2017taiji}
Wen-Rui Hu and Yue-Liang Wu.
\newblock The taiji program in space for gravitational wave physics and the
  nature of gravity, 2017.

\bibitem{luo2016tianqin}
Jun Luo, Li-Sheng Chen, Hui-Zong Duan, Yun-Gui Gong, Shoucun Hu, Jianghui Ji,
  Qi~Liu, Jianwei Mei, Vadim Milyukov, Mikhail Sazhin, et~al.
\newblock Tianqin: a space-borne gravitational wave detector.
\newblock {\em Classical and Quantum Gravity}, 33(3):035010, 2016.

\bibitem{amaro2007intermediate}
Pau Amaro-Seoane, Jonathan~R Gair, Marc Freitag, M~Coleman Miller, Ilya Mandel,
  Curt~J Cutler, and Stanislav Babak.
\newblock Intermediate and extreme mass-ratio inspirals???astrophysics, science
  applications and detection using lisa.
\newblock {\em Classical and Quantum Gravity}, 24(17):R113, 2007.

\bibitem{babak2017science}
Stanislav Babak, Jonathan Gair, Alberto Sesana, Enrico Barausse, Carlos~F
  Sopuerta, Christopher~PL Berry, Emanuele Berti, Pau Amaro-Seoane, Antoine
  Petiteau, and Antoine Klein.
\newblock Science with the space-based interferometer lisa. v. extreme
  mass-ratio inspirals.
\newblock {\em Physical Review D}, 95(10):103012, 2017.

\bibitem{berry2019unique}
Christopher~PL Berry, Scott~A Hughes, Carlos~F Sopuerta, Alvin~JK Chua, Anna
  Heffernan, Kelly Holley-Bockelmann, Deyan~P Mihaylov, M~Coleman Miller, and
  Alberto Sesana.
\newblock The unique potential of extreme mass-ratio inspirals for
  gravitational-wave astronomy.
\newblock {\em arXiv preprint arXiv:1903.03686}, 2019.

\bibitem{gair2013testing}
Jonathan~R Gair, Michele Vallisneri, Shane~L Larson, and John~G Baker.
\newblock Testing general relativity with low-frequency, space-based
  gravitational-wave detectors.
\newblock {\em Living Reviews in Relativity}, 16(1):7, 2013.

\bibitem{barack2004lisa}
Leor Barack and Curt Cutler.
\newblock Lisa capture sources: Approximate waveforms, signal-to-noise ratios,
  and parameter estimation accuracy.
\newblock {\em Physical Review D}, 69(8):082005, 2004.

\bibitem{chua2017augmented}
Alvin~JK Chua, Christopher~J Moore, and Jonathan~R Gair.
\newblock Augmented kludge waveforms for detecting extreme-mass-ratio
  inspirals.
\newblock {\em Physical Review D}, 96(4):044005, 2017.

\bibitem{babak2007kludge}
Stanislav Babak, Hua Fang, Jonathan~R Gair, Kostas Glampedakis, and Scott~A
  Hughes.
\newblock ???kludge??? gravitational waveforms for a test-body orbiting a kerr
  black hole.
\newblock {\em Physical Review D}, 75(2):024005, 2007.

\bibitem{xin2019gravitational}
Shuo Xin, Wen-Biao Han, and Shu-Cheng Yang.
\newblock Gravitational waves from extreme-mass-ratio inspirals using general
  parametrized metrics.
\newblock {\em Physical Review D}, 100(8):084055, 2019.

\bibitem{yunes2010modeling}
Nicolas Yunes, Alessandra Buonanno, Scott~A Hughes, M~Coleman Miller, and
  Yi~Pan.
\newblock Modeling extreme mass ratio inspirals within the effective-one-body
  approach.
\newblock {\em Physical review letters}, 104(9):091102, 2010.

\bibitem{yunes2011extreme}
Nicolas Yunes, Alessandra Buonanno, Scott~A Hughes, Yi~Pan, Enrico Barausse,
  M~Coleman Miller, and William Throwe.
\newblock Extreme mass-ratio inspirals in the effective-one-body approach:
  Quasicircular, equatorial orbits around a spinning black hole.
\newblock {\em Physical Review D}, 83(4):044044, 2011.

\bibitem{han2014gravitational}
Wen-Biao Han.
\newblock Gravitational waves from extreme-mass-ratio inspirals in equatorially
  eccentric orbits.
\newblock {\em International Journal of Modern Physics D}, 23(07):1450064,
  2014.

\bibitem{amaro2018detecting}
Pau Amaro-Seoane.
\newblock Detecting intermediate-mass ratio inspirals from the ground and
  space.
\newblock {\em Physical Review D}, 98(6):063018, 2018.

\bibitem{buonanno1999eff}
Alessandra Buonanno and Thibault Damour.
\newblock Effective one-body approach to general relativistic two-body
  dynamics.
\newblock {\em Physical Review D}, 59(8):084006, 1999.

\bibitem{buonanno2000transition}
Alessandra Buonanno and Thibault Damour.
\newblock Transition from inspiral to plunge in binary black hole coalescences.
\newblock {\em Physical Review D}, 62(6):064015, 2000.

\bibitem{taracchini2014effective}
Andrea Taracchini, Alessandra Buonanno, Yi~Pan, Tanja Hinderer, Michael Boyle,
  Daniel~A Hemberger, Lawrence~E Kidder, Geoffrey Lovelace, Abdul~H Mrou{\'e},
  Harald~P Pfeiffer, et~al.
\newblock Effective-one-body model for black-hole binaries with generic mass
  ratios and spins.
\newblock {\em Physical Review D}, 89(6):061502, 2014.

\bibitem{buonanno2007approaching}
Alessandra Buonanno, Yi~Pan, John~G Baker, Joan Centrella, Bernard~J Kelly,
  Sean~T McWilliams, and James~R van Meter.
\newblock Approaching faithful templates for nonspinning binary black holes
  using the effective-one-body approach.
\newblock {\em Physical Review D}, 76(10):104049, 2007.

\bibitem{purrer2016frequency}
Michael P{\"u}rrer.
\newblock Frequency domain reduced order model of aligned-spin
  effective-one-body waveforms with generic mass ratios and spins.
\newblock {\em Physical Review D}, 93(6):064041, 2016.

\bibitem{husa2016frequency}
Sascha Husa, Sebastian Khan, Mark Hannam, Michael P{\"u}rrer, Frank Ohme,
  Xisco~Jim{\'e}nez Forteza, and Alejandro Boh{\'e}.
\newblock Frequency-domain gravitational waves from nonprecessing black-hole
  binaries. i. new numerical waveforms and anatomy of the signal.
\newblock {\em Physical Review D}, 93(4):044006, 2016.

\bibitem{khan2016frequency}
Sebastian Khan, Sascha Husa, Mark Hannam, Frank Ohme, Michael P{\"u}rrer,
  Xisco~Jim{\'e}nez Forteza, and Alejandro Boh{\'e}.
\newblock Frequency-domain gravitational waves from nonprecessing black-hole
  binaries. ii. a phenomenological model for the advanced detector era.
\newblock {\em Physical Review D}, 93(4):044007, 2016.

\bibitem{chu2016accuracy}
Tony Chu, Heather Fong, Prayush Kumar, Harald~P Pfeiffer, Michael Boyle,
  Daniel~A Hemberger, Lawrence~E Kidder, Mark~A Scheel, and Bela Szilagyi.
\newblock On the accuracy and precision of numerical waveforms: Effect of
  waveform extraction methodology.
\newblock {\em Classical and Quantum Gravity}, 33(16):165001, 2016.

\bibitem{kumar2016accuracy}
Prayush Kumar, Tony Chu, Heather Fong, Harald~P Pfeiffer, Michael Boyle,
  Daniel~A Hemberger, Lawrence~E Kidder, Mark~A Scheel, and Bela Szilagyi.
\newblock Accuracy of binary black hole waveform models for aligned-spin
  binaries.
\newblock {\em Physical Review D}, 93(10):104050, 2016.

\bibitem{pan2014inspiral}
Yi~Pan, Alessandra Buonanno, Andrea Taracchini, Lawrence~E Kidder, Abdul~H
  Mrou{\'e}, Harald~P Pfeiffer, Mark~A Scheel, and B{\'e}la Szil{\'a}gyi.
\newblock Inspiral-merger-ringdown waveforms of spinning, precessing black-hole
  binaries in the effective-one-body formalism.
\newblock {\em Physical Review D}, 89(8):084006, 2014.

\bibitem{hinderer2017foundations}
Tanja Hinderer and Stanislav Babak.
\newblock Foundations of an effective-one-body model for coalescing binaries on
  eccentric orbits.
\newblock {\em Physical Review D}, 96(10):104048, 2017.

\bibitem{cao2017waveform}
Zhoujian Cao and Wen-Biao Han.
\newblock Waveform model for an eccentric binary black hole based on the
  effective-one-body-numerical-relativity formalism.
\newblock {\em Physical Review D}, 96(4):044028, 2017.

\bibitem{Teukolsky}
Saul~A Teukolsky.
\newblock Perturbations of a rotating black hole. 1. fundamental equations for
  gravitational electromagnetic and neutrino field perturbations.
\newblock {\em Astrophys. J.}, 185:635--647, 1973.

\bibitem{han2010gravitational}
Wen-Biao Han.
\newblock Gravitational radiation from a spinning compact object around a
  supermassive kerr black hole in circular orbit.
\newblock {\em Physical Review D}, 82(8):084013, 2010.

\bibitem{han2011constructing}
Wen-Biao Han and Zhoujian Cao.
\newblock Constructing effective one-body dynamics with numerical energy flux
  for intermediate-mass-ratio inspirals.
\newblock {\em Physical Review D}, 84(4):044014, 2011.

\bibitem{han2017excitation}
Wen-Biao Han, Zhoujian Cao, and Yi-Ming Hu.
\newblock Excitation of high frequency voices from intermediate-mass-ratio
  inspirals with large eccentricity.
\newblock {\em Classical and Quantum Gravity}, 34(22):225010, 2017.

\bibitem{cai2016gravitational}
Ronggen Cai, Zhoujian Cao, and Wenbiao Han.
\newblock The gravitational wave models for binary compact objects.
\newblock {\em Chinese Science Bulletin}, 61(14):1525--1535, 2016.

\bibitem{yang2019testing}
Shu-Cheng Yang, Wen-Biao Han, Shuo Xin, and Chen Zhang.
\newblock Testing dispersion of gravitational waves from eccentric
  extreme-mass-ratio inspirals.
\newblock {\em International Journal of Modern Physics D}, 28(15):1950166,
  2019.

\bibitem{cheng2019highly}
Ran CHENG and Wen-biao HAN.
\newblock Highly accurate recalibrate waveforms for extreme-mass-ratio
  inspirals in effective-one-body frames.
\newblock 2019.

\bibitem{barausse2010improved}
Enrico Barausse and Alessandra Buonanno.
\newblock Improved effective-one-body hamiltonian for spinning black-hole
  binaries.
\newblock {\em Physical Review D}, 81(8):084024, 2010.

\bibitem{glampedakis2002zoom}
Kostas Glampedakis and Daniel Kennefick.
\newblock Zoom and whirl: Eccentric equatorial orbits around spinning black
  holes and their evolution under gravitational radiation reaction.
\newblock {\em Physical Review D}, 66(4):044002, 2002.

\bibitem{sago2015calculation}
Norichika Sago and Ryuichi Fujita.
\newblock Calculation of radiation reaction effect on orbital parameters in
  kerr spacetime.
\newblock {\em Progress of Theoretical and Experimental Physics}, 2015(7),
  2015.

\bibitem{gair2006improved}
Jonathan~R Gair and Kostas Glampedakis.
\newblock Improved approximate inspirals of test bodies into kerr black holes.
\newblock {\em Physical Review D}, 73(6):064037, 2006.

\bibitem{finn1992detection}
Lee~S Finn.
\newblock Detection, measurement, and gravitational radiation.
\newblock {\em Physical Review D}, 46(12):5236, 1992.

\bibitem{steinhoff2016Apotential}
Jan Steinhoff, Tanja Hinderer, Alessandra Buonanno, and Andrea Taracchini.
\newblock Dynamical tides in general relativity: Effective action and
  effective-one-body hamiltonian.
\newblock {\em Physical Review D}, 94(10):104028, 2016.

\bibitem{bohe2017improved}
Alejandro Boh{\'e}, Lijing Shao, Andrea Taracchini, Alessandra Buonanno,
  Stanislav Babak, Ian~W Harry, Ian Hinder, Serguei Ossokine, Michael
  P{\"u}rrer, Vivien Raymond, et~al.
\newblock Improved effective-one-body model of spinning, nonprecessing binary
  black holes for the era of gravitational-wave astrophysics with advanced
  detectors.
\newblock {\em Physical Review D}, 95(4):044028, 2017.

\bibitem{sasaki2003analytic}
Misao Sasaki and Hideyuki Tagoshi.
\newblock Analytic black hole perturbation approach to gravitational radiation.
\newblock {\em Living Reviews in Relativity}, 6(1):6, 2003.

\bibitem{fujita2004new}
Ryuichi Fujita and Hideyuki Tagoshi.
\newblock New numerical methods to evaluate homogeneous solutions of the
  teukolsky equation.
\newblock {\em Progress of theoretical physics}, 112(3):415--450, 2004.

\end{thebibliography}

\end{document}